\documentstyle[12pt,graphicx,amsfonts]{article}

\newcommand{\CN}{{\cal N}}

\newcommand{\beq}{\begin{equation}}
\newcommand{\eeq}{\end{equation}}
\newcommand{\bea}{\begin{eqnarray}}
\newcommand{\eea}{\end{eqnarray}}
\newcommand{\ena}{\end{eqnarray}}
\newcommand{\pab}{\Bar{\partial}}
\newcommand{\te}{\q}
\newcommand{\teb}{\Bar \q}
\newcommand{\F}{\Phi}
\newcommand{\Fib}{\Bar{\Phi}}
\newcommand{\Sb}{\Bar{\Sigma}}
\newcommand{\Db}{\Bar{D}}
\def\mathfrak{\bf}
\newcommand {\non}{\nonumber}
\newcommand{\bm}[1]{\mbox{\boldmath$#1$}}
\newcommand{\Ob}{\Bar{\Omega}}
\newcommand{\Tr}{{\rm Tr}}
\renewcommand{\(}{\left(}
\renewcommand{\)}{\right)}
\renewcommand{\[}{\left[}
\renewcommand{\]}{\right]}
\newcommand{\ggl}{\left\{}
\newcommand{\ggr}{\right\}}
\newcommand{\Dcb}{\Bar{\nabla}}
\newcommand{\Dc}{\nabla}
\newcommand{\Lb}{\Bar{\Lambda}}
\newcommand{\Gb}{\Bar{\Gamma}}
\newcommand{\sba}{\Bar{\sigma}}
\newcommand{\DD}{\Delta}
\newcommand{\DDb}{\Bar{\Delta}}

\newcommand{\Xis}{\breve{\Xi}}
\newcommand{\Xib}{\Bar{\Xi}}
\newcommand{\Y}{\Upsilon}
\newcommand{\Yb}{\Bar{\Upsilon}}
\newcommand{\Ys}{\breve{\Upsilon}}
\newcommand{\Qb}{\Bar{Q}}

\def\be{\begin{equation}}
\def\ee{\end{equation}}
\def\bea{\begin{eqnarray}}
\def\eea{\end{eqnarray}}

\def\dt#1{\on{\hbox{\bf .}}{#1}}                
\def\Dot#1{\dt{#1}}
\def\IR{\relax{\rm I\kern-.18em R}}
\def\binomial#1#2{\left(\,{\buildrel
{\raise4pt\hbox{$\displaystyle{#1}$}}\over
{\raise-6pt\hbox{$\displaystyle{#2}$}}}\,\right)}

\def\[{\lfloor{\hskip 0.35pt}\!\!\!\lceil}
\def\]{\rfloor{\hskip 0.35pt}\!\!\!\rceil}

\newcommand{\AmS}{{\protect\the\textfont2
  A\kern-.1667em\lower.5ex\hbox{M}\kern-.125emS}}

\catcode`@=11
\def\un#1{\relax\ifmmode\@@underline#1\else
        $\@@underline{\hbox{#1}}$\relax\fi}
\catcode`@=12

\def\ad{{\kern0.5pt
                   \alpha \kern-5.05pt
\raise5.8pt\hbox{$\textstyle.$}\kern
0.5pt}}

\def\Dot#1{{\kern0.5pt
     {#1} \kern-5.05pt \raise5.8pt\hbox{$\textstyle.$}\kern
0.5pt}}

\def\a{\alpha}
\def\b{\beta}
\def\c{\chi}
\def\d{\delta}
\def\e{\epsilon}

\def\g{\gamma}

\def\i{\iota}

\def\m{\mu}
\def\n{\nu}

\def\q{\theta}
\def\r{\rho}
\def\s{\sigma}

\def\z{\zeta}

\def\F{\Phi}
\def\G{\Gamma}

\def\L{\Lambda}
\def\O{\Omega}

\def\S{\Sigma}

\def\bo{{\raise.15ex\hbox{\large$\Box$}}}               
\def\pa{\partial}                                       
\def\TH{{\raise.2ex\hbox{$\displaystyle \bigodot$}\mskip-4.7mu \llap H
\;}}
\def\face{{\raise.2ex\hbox{$\displaystyle \bigodot$}\mskip-2.2mu \llap
{$\ddot
        \smile$}}}                                      

\def\Bar#1{\overline{#1}}                       
\def\leftrightarrowfill{$\mathsurround=0pt \mathord\leftarrow \mkern-6mu
        \cleaders\hbox{$\mkern-2mu \mathord- \mkern-2mu$}\hfill
        \mkern-6mu \mathord\rightarrow$}
\def\dvec#1{\vbox{\ialign{##\crcr
        \leftrightarrowfill\crcr\noalign{\kern-1pt\nointerlineskip}
        $\hfil\displaystyle{#1}\hfil$\crcr}}}           
\def\dt#1{{\buildrel {\hbox{\LARGE .}} \over {#1}}}     

\def\frac#1#2{{\textstyle{#1\over\vphantom2\smash{\raise.20ex
        \hbox{$\scriptstyle{#2}$}}}}}                   
\def\sfrac#1#2{{\vphantom1\smash{\lower.5ex\hbox{\small$#1$}}\over
        \vphantom1\smash{\raise.4ex\hbox{\small$#2$}}}} 
\def\bfrac#1#2{{\vphantom1\smash{\lower.5ex\hbox{$#1$}}\over
        \vphantom1\smash{\raise.3ex\hbox{$#2$}}}}       
\def\afrac#1#2{{\vphantom1\smash{\lower.5ex\hbox{$#1$}}\over#2}}    
\def\on#1#2{\mathop{\null#2}\limits^{#1}}               

\newskip\humongous \humongous=0pt plus 1000pt minus 1000pt
\def\caja{\mathsurround=0pt}
\def\eqalign#1{\,\vcenter{\openup2\jot \caja
        \ialign{\strut \hfil$\displaystyle{##}$&$
        \displaystyle{{}##}$\hfil\crcr#1\crcr}}\,}
\newif\ifdtup

  \def\pp{{\mathchoice
          {
              \kern 1pt%
              \raise 1pt
              \vbox{\hrule width5pt height0.4pt depth0pt
                    \kern -2pt
                    \hbox{\kern 2.3pt
                          \vrule width0.4pt height6pt depth0pt
                          }
                    \kern -2pt
                    \hrule width5pt height0.4pt depth0pt}%
                    \kern 1pt
           }
            {
              \kern 1pt%
              \raise 1pt
              \vbox{\hrule width4.3pt height0.4pt depth0pt
                    \kern -1.8pt
                    \hbox{\kern 1.95pt
                          \vrule width0.4pt height5.4pt depth0pt
                          }
                    \kern -1.8pt
                    \hrule width4.3pt height0.4pt depth0pt}%
                    \kern 1pt
            }
            {
              \kern 0.5pt%
              \raise 1pt
              \vbox{\hrule width4.0pt height0.3pt depth0pt
                    \kern -1.9pt  
                    \hbox{\kern 1.85pt
                          \vrule width0.3pt height5.7pt depth0pt
                          }
                    \kern -1.9pt
                    \hrule width4.0pt height0.3pt depth0pt}%
                    \kern 0.5pt
            }
            {
              \kern 0.5pt%
              \raise 1pt
              \vbox{\hrule width3.6pt height0.3pt depth0pt
                    \kern -1.5pt
                    \hbox{\kern 1.65pt
                          \vrule width0.3pt height4.5pt depth0pt
                          }
                    \kern -1.5pt
                    \hrule width3.6pt height0.3pt depth0pt}%
                    \kern 0.5pt
            }
        }}

  \def\mm{{\mathchoice
                       {
                             \kern 1pt
               \raise 1pt    \vbox{\hrule width5pt height0.4pt depth0pt
                                  \kern 2pt
                                  \hrule width5pt height0.4pt depth0pt}
                             \kern 1pt}
                       {
                            \kern 1pt
               \raise 1pt \vbox{\hrule width4.3pt height0.4pt depth0pt
                                  \kern 1.8pt
                                  \hrule width4.3pt height0.4pt depth0pt}
                             \kern 1pt}
                       {
                            \kern 0.5pt
               \raise 1pt
                            \vbox{\hrule width4.0pt height0.3pt depth0pt
                                  \kern 1.9pt
                                  \hrule width4.0pt height0.3pt depth0pt}
                            \kern 1pt}
                       {
                           \kern 0.5pt
             \raise 1pt  \vbox{\hrule width3.6pt height0.3pt depth0pt
                                  \kern 1.5pt
                                  \hrule width3.6pt height0.3pt depth0pt}
                           \kern 0.5pt}
                       }}

\def\pd{{\kern0.5pt
                   + \kern-5.05pt \raise5.8pt\hbox{$\textstyle.$}\kern
0.5pt}}

\def\pmd{{\kern0.5pt
                  \pm \kern-5.05pt \raise6.3pt\hbox{$\textstyle.$}\kern1.5pt}}

\def\md{{\mathchoice
   {
      {{\kern 1pt - \kern-6.2pt \raise5pt\hbox{$\textstyle.$}\kern 1pt}}}
    {
      {{\kern 1pt - \kern-6.2pt \raise5pt\hbox{$\textstyle.$}\kern 1pt}}}
    {
      {\kern0.5pt - \kern-5.05pt \raise3.4pt\hbox{$\textstyle.$}\kern0.5pt}}
    {
      {\kern0.5pt - \kern-5.05pt \raise3.4pt\hbox{$\textstyle.$}\kern0.5pt}}}}

\def\ad{{\dot{\alpha}}}
\def\bd{{\dot{\beta}}}

\def\pp{{\mathchoice
          {
              \kern 1pt%
              \raise 1pt
              \vbox{\hrule width5pt height0.4pt depth0pt
                    \kern -2pt
                    \hbox{\kern 2.3pt
                          \vrule width0.4pt height6pt depth0pt
                          }
                    \kern -2pt
                    \hrule width5pt height0.4pt depth0pt}%
                    \kern 1pt
           }
            {
              \kern 1pt%
              \raise 1pt
              \vbox{\hrule width4.3pt height0.4pt depth0pt
                    \kern -1.8pt
                    \hbox{\kern 1.95pt
                          \vrule width0.4pt height5.4pt depth0pt
                          }
                    \kern -1.8pt
                    \hrule width4.3pt height0.4pt depth0pt}%
                    \kern 1pt
            }
            {
              \kern 0.5pt%
              \raise 1pt
              \vbox{\hrule width4.0pt height0.3pt depth0pt
                    \kern -1.9pt  
                    \hbox{\kern 1.85pt
                          \vrule width0.3pt height5.7pt depth0pt
                          }
                    \kern -1.9pt
                    \hrule width4.0pt height0.3pt depth0pt}%
                    \kern 0.5pt
            }
            {
              \kern 0.5pt%
              \raise 1pt
              \vbox{\hrule width3.6pt height0.3pt depth0pt
                    \kern -1.5pt
                    \hbox{\kern 1.65pt
                          \vrule width0.3pt height4.5pt depth0pt
                          }
                    \kern -1.5pt
                    \hrule width3.6pt height0.3pt depth0pt}%
                    \kern 0.5pt
            }
        }}

  \def\mm{{\mathchoice
                       {
                             \kern 1pt
               \raise 1pt    \vbox{\hrule width5pt height0.4pt depth0pt
                                  \kern 2pt
                                  \hrule width5pt height0.4pt depth0pt}
                             \kern 1pt}
                       {
                            \kern 1pt
               \raise 1pt \vbox{\hrule width4.3pt height0.4pt depth0pt
                                  \kern 1.8pt
                                  \hrule width4.3pt height0.4pt depth0pt}
                             \kern 1pt}
                       {
                            \kern 0.5pt
               \raise 1pt
                            \vbox{\hrule width4.0pt height0.3pt depth0pt
                                  \kern 1.9pt
                                  \hrule width4.0pt height0.3pt depth0pt}
                            \kern 1pt}
                       {
                           \kern 0.5pt
             \raise 1pt  \vbox{\hrule width3.6pt height0.3pt depth0pt
                                  \kern 1.5pt
                                  \hrule width3.6pt height0.3pt depth0pt}
                           \kern 0.5pt}
                       }}

\def\pd{{\kern0.5pt
                   + \kern-5.05pt \raise5.8pt\hbox{$\textstyle.$}\kern
0.5pt}}

\def\pmd{{\kern0.5pt
                  \pm \kern-5.05pt \raise6.3pt\hbox{$\textstyle.$}\kern1.5pt}}

\def\md{{\mathchoice
   {
      {{\kern 1pt - \kern-6.2pt \raise5pt\hbox{$\textstyle.$}\kern 1pt}}}
    {
      {{\kern 1pt - \kern-6.2pt \raise5pt\hbox{$\textstyle.$}\kern 1pt}}}
    {
      {\kern0.5pt - \kern-5.05pt \raise3.4pt\hbox{$\textstyle.$}\kern0.5pt}}
    {
      {\kern0.5pt - \kern-5.05pt \raise3.4pt\hbox{$\textstyle.$}\kern0.5pt}}}}

\def\dslash{\not{\hbox{\kern-2pt $\partial$}}}
\def\Dslash{\not{\hbox{\kern-4pt $D$}}}
\def\pslash{\not{\hbox{\kern-2.3pt $p$}}}
 \newtoks\slashfraction
 \slashfraction={.13}
 \def\slash#1{\setbox0\hbox{$ #1 $}
 \setbox0\hbox to \the\slashfraction\wd0{\hss \box0}/\box0 }

\font\ro=cmsy10                          
\def\kcr{{\hbox{\ro \char'170}}}                
\def\ktl{{\hbox{\ro \char'170}}}        
\def\ktr{{\hbox{\ro \char'170}}}        
\def\kbl{{\hbox{\ro \char'170}}}        
\def\kbr{{\hbox{\ro \char'170}}}        

\def\plpl{\raise-2pt\hbox{$\raise3pt\hbox{$_+$}\hskip-6.67pt\raise0.0pt
\hbox{$^+$}\hskip 0.01pt$}}
\def\mimi{\raise-2pt\hbox{$\raise3pt\hbox{$_-$}\hskip-6.67pt\raise0.0pt
\hbox{$^-$}\hskip 0.01pt$}}

\def\bo{{\raise.15ex\hbox{\large$\Box$}}}               
\def\pa{\partial}                                       
\def\TH{{\raise.2ex\hbox{$\displaystyle \bigodot$}\mskip-4.7mu \llap H \;}}
\def\face{{\raise.2ex\hbox{$\displaystyle \bigodot$}\mskip-2.2mu \llap {$\ddot
        \smile$}}}                                      

\def\Bar#1{\overline{#1}}                       
\def\leftrightarrowfill{$\mathsurround=0pt \mathord\leftarrow \mkern-6mu
        \cleaders\hbox{$\mkern-2mu \mathord- \mkern-2mu$}\hfill
        \mkern-6mu \mathord\rightarrow$}
\def\dvec#1{\vbox{\ialign{##\crcr
        \leftrightarrowfill\crcr\noalign{\kern-1pt\nointerlineskip}
        $\hfil\displaystyle{#1}\hfil$\crcr}}}           
\def\dt#1{{\buildrel {\hbox{\LARGE .}} \over {#1}}}     

\def\frac#1#2{{\textstyle{#1\over\vphantom2\smash{\raise.20ex
        \hbox{$\scriptstyle{#2}$}}}}}                   
\def\sfrac#1#2{{\vphantom1\smash{\lower.5ex\hbox{\small$#1$}}\over
        \vphantom1\smash{\raise.4ex\hbox{\small$#2$}}}} 
\def\bfrac#1#2{{\vphantom1\smash{\lower.5ex\hbox{$#1$}}\over
        \vphantom1\smash{\raise.3ex\hbox{$#2$}}}}       
\def\afrac#1#2{{\vphantom1\smash{\lower.5ex\hbox{$#1$}}\over#2}}    
\def\on#1#2{\mathop{\null#2}\limits^{#1}}               

\parskip=\medskipamount                 
\thicklines                         

\def\oldheadpic{                                
        \setlength{\unitlength}{.4mm}
        \thinlines
        \par
        \begin{picture}(349,16)
        \put(325,16){\line(1,0){4}}
        \put(330,16){\line(1,0){4}}
        \put(340,16){\line(1,0){4}}
        \put(335,0){\line(1,0){4}}
        \put(340,0){\line(1,0){4}}
        \put(345,0){\line(1,0){4}}
        \put(329,0){\line(0,1){16}}
        \put(330,0){\line(0,1){16}}
        \put(339,0){\line(0,1){16}}
        \put(340,0){\line(0,1){16}}
        \put(344,0){\line(0,1){16}}
        \put(345,0){\line(0,1){16}}
        \put(329,16){\oval(8,32)[bl]}
        \put(330,16){\oval(8,32)[br]}
        \put(339,0){\oval(8,32)[tl]}
        \put(345,0){\oval(8,32)[tr]}
        \end{picture}
        \par
        \thicklines
        \vskip.2in}
\def\oldtitle#1#2#3#4{\oldheadpic\begin{center}\vglue.5in{\large\bf #1}\\[.6in]
        {#2}\\[.1in] {\it Department of Physics and Astronomy}\\
        {\it University of Maryland, College Park, MD 20742}\\[.6in]
        Physics Publication \#{#3}\\ {#4}\\[1.5in] {\bf ABSTRACT}\\[.1in]
        \end{center} \begin{quotation}}                 
\def\oldTitle#1#2#3#4#5#6#7{\oldheadpic\begin{center} \vglue .4in
        {\large\bf #1}\\[.4in]
        {#2}\\[.1in] {\it Department of Physics and Astronomy}\\
        {\it University of Maryland, College Park, MD 20742}\\[.1in]
        {#3}\\[.1in] {\it {#4}}\\ {\it {#5}}\\[.4in]
        Physics Publication \#{#6}\\ {#7}\\[.5in] {\bf ABSTRACT}\\[.1in]
        \end{center} \begin{quotation}}                 
\def\border{                                            
        \setlength{\unitlength}{1mm}
        \newcount\xco
        \newcount\yco
        \xco=-21
        \yco=12
        \begin{picture}(140,0)
        \put(\xco,\yco){$\ktl$}
        \advance\yco by-1
        {\loop
        \put(\xco,\yco){$\kcr$}
        \advance\yco by-2
        \ifnum\yco>-240
        \repeat
        \put(\xco,\yco){$\kbl$}}
        \xco=158
        \yco=12
        \put(\xco,\yco){$\ktr$}
        \advance\yco by-1
        {\loop
        \put(\xco,\yco){$\kcr$}
        \advance\yco by-2
        \ifnum\yco>-240
        \repeat
        \put(\xco,\yco){$\kbr$}}
        \put(-20,13){\tiny **University of Maryland * Center for String and
         Particle  Theory* Physics Department***University of Maryland *Center
        for String and Particle  Theory** }
        \put(-20,-241.5){\tiny **University of Maryland * Center for String and
         Particle  Theory* Physics Department***University of Maryland *Center
        for String and Particle  Theory** }
        \end{picture}
        \par\vskip-8mm}
\def\bordero{                                           
        \setlength{\unitlength}{1mm}
        \newcount\xco
        \newcount\yco
        \xco=-31
        \yco=12
        \begin{picture}(140,0)
        \put(\xco,\yco){$\ktl$}
        \advance\yco by-1
        {\loop
        \put(\xco,\yco){$\kclr$}
        \advance\yco by-2
        \ifnum\yco>-240
        \repeat
        \put(\xco,\yco){$\kbl$}}
        \xco=151
        \yco=12
        \put(\xco,\yco){$\ktr$}
        \advance\yco by-1
        {\loop
        \put(\xco,\yco){$\kcr$}
        \advance\yco by-2
        \ifnum\yco>-240
        \repeat
        \put(\xco,\yco){$\kbr$}}
        \put(-20,12){\ooo bacdefghidfghghdhededbihdgdfdfhhdheidhdhebaaahjhhdahba

hgdedge
   hgfdiehhgdigicba}
        \put(-20,-241.5){\ooo ababaighefdbfghgeahgdfgafagihdidihiidhiagfedhadbfd

ecdcdfa
   gdcbhaddhbgfchbgfdacfediacbabab}
        \end{picture}
        \par\vskip-8mm}
\def\headpic{                                           
        \indent
        \setlength{\unitlength}{.4mm}
        \thinlines
        \par
        \begin{picture}(29,16)
        \put(165,16){\line(1,0){4}}
        \put(170,16){\line(1,0){4}}
        \put(180,16){\line(1,0){4}}
        \put(175,0){\line(1,0){4}}
        \put(180,0){\line(1,0){4}}
        \put(185,0){\line(1,0){4}}
        \put(169,0){\line(0,1){16}}
        \put(170,0){\line(0,1){16}}
        \put(179,0){\line(0,1){16}}
        \put(180,0){\line(0,1){16}}
        \put(184,0){\line(0,1){16}}
        \put(185,0){\line(0,1){16}}
        \put(169,16){\oval(8,32)[bl]}
        \put(170,16){\oval(8,32)[br]}
        \put(179,0){\oval(8,32)[tl]}
        \put(185,0){\oval(8,32)[tr]}
        \end{picture}
        \par\vskip-6.5mm
        \thicklines}
\def\title#1#2#3#4{\border\headpic {\hbox to\hsize{#4 \hfill UMDEPP #3}}\par
        \begin{center} \vglue .5in {\large\bf #1}\\[.6in]
        {#2}\\[.1in] {\it Department of Physics and Astronomy}\\
        {\it University of Maryland, College Park, MD 20742}\\[1.5in]
        {\bf ABSTRACT}\\[.1in] \end{center} \begin{quotation}}  
\def\Title#1#2#3#4#5#6#7{\border\headpic
        {\hbox to\hsize{#7 \hfill UMDEPP #6}}\par
        \begin{center} \vglue .4in {\large\bf #1}\\[.4in]
        {#2}\\[.1in] {\it Department of Physics and Astronomy}\\
        {\it University of Maryland, College Park, MD 20742}\\[.1in]
        {#3}\\[.1in] {\it {#4}}\\ {\it {#5}}\\[.5in] {\bf ABSTRACT}\\[.1in]
        \end{center} \begin{quotation}}                 
\def\endtitle{\end{quotation}\newpage}                  

\def\qd{{\kern0.5pt
                   q \kern-5.05pt \raise5.8pt\hbox{$\textstyle.$}\kern
0.5pt}}

\begin{document}

\def\dt#1{\on{\hbox{\bf .}}{#1}}                
\def\Dot#1{\dt{#1}}

\def\gfrac#1#2{\frac {\scriptstyle{#1}}
        {\mbox{\raisebox{-.6ex}{$\scriptstyle{#2}$}}}}
\def\gg{{\hbox{\sc g}}}
\border\headpic {\hbox to\hsize{August 2005 \hfill
{UMDEPP 05-134}}}
\par
{$~$ \hfill
{Bicocca--FT--05--21}}
\par
~~~
{$~$ \hfill {hep-th/0508187}}
\par

\setlength{\oddsidemargin}{0.3in}
\setlength{\evensidemargin}{-0.3in}
\begin{center}
\vglue .10in
{\large\bf 6D Supersymmetry, Projective Superspace \&\\
4D, $\bm {\cal N}$ $\bm =$ 1 Superfields\footnote
{Supported in part  by National Science Foundation Grant
PHY-0354401, INFN, PRIN prot. \newline $~~~\,~~$ $2003023852\_008$ and
the European Commission RTN program MRTN--CT--2004--005104.}\  }
\\[.35in]

S.\, James Gates, Jr.${}^\dag$\footnote{gatess@wam.umd.edu},
Silvia Penati${}^\star$\footnote{Silvia.Penati@mib.infn.it} and
Gabriele Tartaglino-Mazzucchelli${}^\star$\footnote{Gabriele.Tartaglino@mib.infn.it}
\\[0.3in]
${}^\dag${\it Center for String and Particle Theory\\
Department of Physics, University of Maryland\\
College Park, MD 20742-4111 USA}\\[0.1in]
{\it {and}}\\[0.1in]
${}^\star${\it Dipartimento di Fisica, Universit\`a degli studi
Milano-Bicocca\\and INFN, Sezione di Milano, piazza della Scienza 3,
I-20126 Milano, Italy}\\[0.6in]

{\bf ABSTRACT}\\[.01in]
\end{center}
\begin{quotation}
{In this note, we establish the formulation of 6D, $\CN$ $=$ $1$
hypermultiplets in terms of 4D chiral-nonminimal (CNM) scalar
multiplets. The coupling of these to 6D, $\CN$ $=$ $1$ Yang-Mills
multiplets is described. A 6D, $\CN$ $=$ $1$ projective superspace
formulation is given in which the above multiplets
naturally emerge. The covariant superspace quantization of these
multiplets is studied in details.}

${~~~}$ \newline ${~~~}$ \newline PACS: 04.65.+e, 11.15.-q, 11.30.Pb, 12.60.Jv

\endtitle

\setcounter{equation}{0}
\section{Introduction}

~~~~Six dimensions is the highest one in which supersymmetric multiplets 
possessing states of maximum helicity one-half exist.  The 6D, $\cal N$ $=$ $1$
hypermultiplet can be subjected to  dimensional reduction to obtain a 5D, 
$\cal N$ $=$ $1$ hypermultiplet, a 4D, $\cal N$ $=$ $2$ hypermultiplet, a 3D, 
$\cal N$ $=$ $4$ hypermultiplet, a 2D, $\cal N$ $=$ $4$ hypermultiplet and a 
1D, $\cal N$ $=$ $8$ hypermultiplet.   It is known in these lower dimensions 
additional hypermultiplets unrelated to this chain of reductions also appear.  
These other hypermultiplets are by definition ``twisted'' versions of the 
`standard' hypermultiplet that descends from six dimensions. In fact, it has 
been conjectured \cite{GK1} the number $N_{\rm {H. \, M.}}(D)$ of
{\it {distinct}} on-shell hypermultiplet representations in a spacetime with 
bosonic dimension $D$ obeys the  rule
\be
N_{\rm {H. \, M.}}(D) ~=~ 2^{5 \, - \, D}  ~~
\ee
for $D$ $\le$ $5$ and by definition $N_{\rm {H. \, M.}}(6)$ $=$ $1$.  The 
origin of this formula is at present not understood.  However, there has in 
more recent times begun to emerge evidence this is related to the 
representation theory of certain Clifford algebras and K-theory \cite{GLP}.

So the study of the 6D, $\cal N$ $=$ $1$ hypermultiplet has diverse 
applications in many contexts.  In the following, we shall show there are two 
possible 4D, $\cal N$ $=$ $1$ formulations for the 6D hypermultiplets.  One of 
these involves using a pair of chiral superfields (the CC formulation, for 
example was used in the work of \cite{ArkaniHamedTB}) while the other involves 
one chiral superfield and one complex linear superfield (the CNM formulation). 
These two formulations are related by a duality transformation.

As long as one is concentrated only upon classical considerations
(with the possible exception on issues of the vacuum state) both
formulations are equivalent.  However, if applications involve
quantum mechanical considerations, there are certain advantages of
the CNM formulation.  In the CC formulation it is possible to use
4D, $\cal N$ $=$ $1$ supergraph techniques.  In the CNM
formulation this can be augmented by the use of projective
supergraph techniques 
\cite{ProjectiveSuperspace,ProjectiveSuperspaceGR,ProjectiveSuperspaceK}.
The point is projective supergraphs
are more efficient calculational techniques that manifest more
possible cancellations between different fields than can be seen
with the use of the supergraphs associated with ordinary
superspace. Concerning $\cal N$ = 2 covariance, an alternative
approach would be use of the powerful harmonic
superspace [7].  Classically, the two approaches are
closely linked, but there is concrete evidence that
only the quantum calculations in projective superspace
reduce naturally to those of $\cal N$ = 1 superspace.
Thus in any intricate quantum calculation, the CNM
approach seems most likely the superior one to employ which can be 
directly compared to $\CN=1$ computations. 
We also have focused our attention only on projective superspace instead
of harmonic superspace as this insure the absence of harmonic
singularities \cite{HarmSing} known to occur at one and two loops.
Although these have been resolved, it seems likely order-by-order
this must be implemented to have any ambiguity \cite{Kuz} removed.

With this in mind, a primary purpose of this work is to carry out
the quantization of the CNM formulation of the 6D $\cal N$ $=$ $1$
hypermultiplet.  As is well known, the CNM pair possesses a
natural extension to projective superspace.  Thus as part of our
presentation we include a discussion of the implications of the
projective superspace for the present considerations.  The paper
is organized as follows.

Using the old suggestion by Siegel \cite{siegel}, the first section
describes the use of ordinary 4D $\CN$ $=$ 1 superspace to
describe free 6D $\CN$ $=$ 1 hypermultiplets.  Both CC and
CNM descriptions are given for the free hypermultiplet.  The 4D
formalism is shown capable of describing both 6D (1,0) and (0,1)
Weyl spinors which can occur within hypermultiplets.  Duality
transformations are shown to exist between the 6D theories,
in complete analogy to the lower dimensional cases where a
$\CN$ $=$ 1 ordinary superspace formalism can be used.

The next section describes the coupling to the non-Abelian Yang-Mills 
supermultiplet. This is carried out for both CC and CNM systems.  The 
superspace geometry of the vector multiplets is given and the connections for 
all directions of the six dimensional manifold are shown.  It is  again shown 
the 4D formalism is capable of describing both 6D (1,0) and (0,1) Weyl spinors 
which can occur within vector multiplets.  A restriction is observed to require
a vector multiplet containing one 6D Weyl spinor can {\em only} be coupled to a
hypermultiplet containing the opposite 6D Weyl spinor.

The following section discusses the quantization of the 6D hypermultiplet and 
vector multiplet.  It is shown these can be constructed by previous known 
procedures which have been applied to four dimensions.  In the CNM case, an 
infinite tower of ghosts are found to decouple.

The subsequent section addresses the problem of embedding the 6D
CNM description of the hypermultiplet and vector
multiplet in a projective superspace. This is achieved and it
is shown the previously known structures from the case of 4D
remain intact for 6D.  Tropical and polar multiplets are found.
The embedding of the coupled hypermultiplet/Yang-Mills
multiplet system is presented.  The quantization of the polar
multiplet is discussed and its propagator inferred by exploiting the
analogies with the 4D case.

We close with a set of concluding remarks, perspectives and include
an appendix explaining our notations and conventions with
regard to six dimensions.

\setcounter{equation}{0}
\section{6D, $\bm {\cal N}$ $\bm =$ $\bm 1$ Hypermultiplets}

~~~~In this section we discuss two alternative partially on--shell
descriptions of the six--dimensional $\CN=1$ hypermultiplet
\cite{SUSYKHSTK6d} obtained by using chiral--chiral (CC) and
chiral--nonminimal (CNM) multiplets, whose definitions are
inspired by their four--dimensional analogues.  Following closely
the approach to $10$D $\CN=1$ SYM of \cite{MaSaSi},  the CC
formulation has been already introduced \cite{ArkaniHamedTB}.
A formulation in terms of chiral--nonminimal multiplets (CNM) is
proposed in this note as a new contribution to the literature.

We remind our readers in six dimensions the  physical degrees
of freedom of a $\CN=1$ hypermultiplet are described
by two complex scalars and a 6D Weyl spinor \cite{SUSYKHSTK6d}.
Since in six dimensions there are two different kinds of
Weyl spinors, we define $\CN=(1,0)$ and $\CN=(0,1)$ hypermultiplets,
depending on the 6D chirality of the Weyl spinor (we refer to the appendix
for the spinor notations we adopt).
An on--shell description in terms of 4D multiplets must correctly
describe the six dimensional dynamics of these degrees of freedom.

Following the approach used previously  \cite{ArkaniHamedTB,MaSaSi}
 to describe 10D $\CN=1$ SYM and theories in
D--dimensions with $5\leq {\rm D} \leq 10$, we use a formalism
explicitly covariant under 4D supersymmetry. We parametrize
the six--dimensional spacetime by bosonic coordinates $x_i$, $i=0,
\cdots , 5$. The first four $x_i$, together with the grassmannian
coordinates $(\q_\a,{\Bar \q}_\ad)$ describe the ordinary
four--dimensional $\CN=1$ superspace (we use 4D notations and
conventions of \cite{SUPERSPACE}). Defining the complex
coordinates
\be
\eqalign{ &z\equiv{1\over 2}(x_4+ix_5)~~~~~,~~~~~
\pa\equiv{\pa\over \pa z}=\pa_4-i\pa_5~~~~;~~~~  \cr
&\Bar{z}\equiv{1\over 2}(x_4-ix_5)~~~~~,~~~~~ \pab\equiv{\pa\over
\pa \Bar{z}}=\pa_4+i\pa_5~~~~,~~~~ }
\label{defZbZ}
\ee
the
algebra of supercovariant derivatives is the ordinary one $\{ D_\a
, \bar{D}_\ad \} = i (\s^a)_{\a\ad} \pa_a$, $a = 0,1,2,3$
supplemented by the extra conditions $[ D_\a , \pa ] = [
\bar{D}_\ad , \pa ] = [ D_\a , \pab ] = [ \bar{D}_\ad , \pab ]
=  [ \pa , \pab ]  = 0
$.

The on--shell description of the 6D hypermultiplet in terms of
chiral multiplets \cite{ArkaniHamedTB} is accomplished by the introduction
of two chiral superfields $\O_{\pm}(x_i, \q_\a, \bar{\q}_\ad)$
\beq
\Db_\ad \Omega_\pm=0~~~,~~~D_\a\Bar{\Omega}_\pm=0~~~,
\label{defkinCC}
\eeq
which realize linearly the 4D, $\CN = 1$ supersymmetry and whose components
are functions of the six bosonic coordinates.
The action
\be
\eqalign{
S_{CC}~=~&\int d^6x\, d^4\q \, \Big[~\Bar{\Omega}_+ \, \Omega_+ ~+~
\Bar{\Omega}_- \, \Omega_-~\Big] ~+~
\int d^6x \, d^2\te \Big[~\Omega_+\, \pa \,\Omega_-~\Big] \cr
&+~\int d^6x\, d^2\teb \Big[~\Bar{\Omega}_+\,\pab \,\Bar{\Omega}_-~\Big]~,
}
\label{SCC}
\eeq
when reduced to components and with auxiliary fields integrated out
by the use of their algebraic equations of motion describes correctly the
free propagation of the physical degrees of freedom of the $(1,0)$
hypermultiplet.  In fact, defining component fields via
\bea
&A_\pm=\Omega_\pm|~~~,~~~\psi_\pm^\a=D^\a\Omega_\pm|~~~,~~~
F_\pm=D^2\Omega_\pm|~~~,
\eea
and eliminating the auxiliary fields from the action (\ref{SCC}), we obtain
\beq
S^0_{CC}=\int d^6x \Big[~\Bar{A}_+\Box_6 A_++\Bar{A}_-\Box_6 A_-
-\Bar{\psi}_+^\ad i\pa_{\a \ad}\psi^\a_+
-\Bar{\psi}_-^\ad i\pa_{\a \ad}\psi^\a_-
-\psi_-^\a\pa\,\psi_{+\a}-\Bar{\psi}^\ad_-\pab\,\Bar{\psi}_{+\ad}~\Big]~,~~~
\label{S0CC}
\eeq
where $\Box_6 \equiv \pa^\mu \pa_\mu =  \Box_4 + \pa\pab$ is
the D'Alambertian operator in six dimensions.

This action describes the free dynamics of two complex scalars
$A_{\pm}$ and a 6D Weyl spinor $(\psi^\a_+ \, ,~\Bar{\psi}^\ad_-)$, as
can easily be inferred by comparing the fermionic part of
(\ref{S0CC}) with the action (\ref{free6d(1,0)}) in the appendix
for a free 6D $(1,0)$ Weyl spinor.

An alternative description can be given in terms of a CNM multiplet.
To this end, we introduce a pair of superfields $(\Phi, \S)$ whose
covariant definitions are inspired by the four dimensional chiral and
complex linear superfields \cite{SUPERSPACE,CNM,CNMmassless}
\beq
\Db_\ad\F=0~~~,~~~D_\a\Fib=0~~~,~~~ \Db^2\S=\pa\,\F~~~,~~~
D^2\Sb=\pab\,\Fib~~~,
\label{defkinCNM}
\eeq
In analogy with the four dimensional case we define the components fields as
\bea
&A=\F|~~~,~~~\psi_\a=D_\a\F|~~~,~~~F=D^2\F|~~~,\non\\
&B=\S|~~~,~~~\Bar{\z}_\ad=\Db_\ad\S|~~~,~~~H=D^2\S|~~~,\\
&\rho_\a=D_\a\S|~~~,~~~p_{\a\ad}=\Db_\ad D_\a\S|~~~,~~~
\Bar{\b}_\ad={1\over 2}D^\a\Db_\ad D_\a\S|~~~.\non
\eea
These are functions of the 6D spacetime coordinates.

The action describing the free propagation of these superfields is
\beq
S_{CNM}=\int d^6xd^4\q \Big[~\Fib\F-\Sb\S~\Big]~.
\label{SCNM}
\eeq
Due to the constraints (\ref{defkinCNM}), when reduced to components it
takes the form
\be
\eqalign{
S_{CNM}~=&\int d^6x \, \Big[~\Bar{A}\Box_4 A ~+~ \Bar{B}\Box_4 B
~-~ i\,\Bar{\psi}^\ad \pa_{\a \ad}\psi^\a ~-~ i \,\Bar{\zeta}^\ad  \pa_{\a \ad}
 \zeta^\a  \cr
&~~~~~~~~\,~ ~-~\z^\a\pa \, \psi_\a ~-~\Bar{\z}^\ad\pab\, \Bar{\psi}
_\ad  ~+~ \Bar{A}\, \pab\pa\,  A ~-~ B\,\pab\, \Bar{F} ~-~ \Bar{B}\,\pa\,
F ~~  \cr
&~~~~\,~~~~~ ~+~ \Bar{F}F ~-~\Bar{H} H
~+~ \beta^\a \rho_\a ~+~ \Bar{\rho}^\ad \Bar{\b}_\ad ~-~
\Bar{p}^{\a \ad}p_{\a \ad}~\Big]~.~~~}
\label{SCNMcomp}
\ee
The
auxiliary fields $F, H, \b_\a, \r_\a, p_{\a\ad}$ and their
hermitian conjugates satisfy algebraic equations of motion and can
be eliminated. The result is
\beq
S^0_{CNM}=\int d^6x
\Big[~\Bar{B}\Box_6 B \,+\, \Bar{A}\Box_6 A \,-\, \Bar{\psi}^\ad
i\pa_{\a \ad}\psi^\a \,-\, \Bar{\zeta}^\ad i \pa_{\a \ad} \zeta^\a
 \,-\, \z^\a\pa\,\psi_\a \,-\,
\Bar{\z}^\ad\pab\,\Bar{\psi}_\ad~\Big]~.~~~~~~~~
\label{S0CNM}
\eeq
Here again we see the free dynamics of a 6D $\CN=(1,0)$ massless
hypermultiplet which has as physical degrees of freedom the two complex
scalars $A,B$ and the $(1,0)$ 6D Weyl spinor $(\psi^\a \, ,~~ \Bar{\z}^\ad) $.
Our results in (\ref{defkinCNM}-\ref{S0CNM}) can also be seen to be a 
dimensional oxidation of the recent CNM description \cite{KuzenkoLinch} of the 
5D, $\cal N$ $=$ 1 hypermultiplet.

There are some interesting and subtle differences in the two formulations.
To reach (\ref{S0CC}) from (\ref{SCC}) both auxiliary fields $F_+$ and
$F_-$ were removed via their equations of motion.  The result of this is
to insure the six dimensional D'Alambertian operator appears for
$A_{\pm}$ in (\ref{SCC}).  Something rather different occurs in deriving
 (\ref{S0CNM}) from (\ref{SCNMcomp}).  The six dimensional D'Alambertian
 is already present for $A$ even prior to the elimination of any auxiliary
 field.

A well--known fact in four dimensions is  chiral and complex--linear
massless superfields are dual to each other \cite{SUPERSPACE}. The CC
and CNM 6D massless hypermultiplets introduced above are the analogues
of the 4D chiral and complex--linear massless superfields, 
respectively\footnote{Note the constraint for the nonminimal multiplet $\S$ as 
given in (\ref{defkinCNM}) is modified respect \newline $~~\,~~~$ to the 
ordinary $\bar{D}^2 \S = 0$, in analogy to the 4D CNM generalizations
proposed in \cite{CNM} and \newline $~~\,~~~$ further studied in 
\cite{CNM0404222}.
However, as discussed in \cite{ProjectiveSuperspaceGR},
the duality properties between a pair \newline $~~~\,~~$  of massive chirals 
and a pair of a chiral and a nonminimal superfield survive the more  \newline 
$~~\,~~~$ general case $\Db^2\S=m\F$.}.
We then expect the 4D duality of the two multiplets
to be extended to the 6D hypermultiplets.

This is easily implemented by introducing the following action
\beq
\int d^6xd^4\te\Big[\Bar{\Psi}\Psi-\Bar{\eta}\eta+
Y(\Db^2\eta-\pa\Psi)+\Bar{Y}(D^2\Bar{\eta}-\pab\,\Bar{\Psi})\Big]~,
\label{CCNMdual}
\eeq
where $\Psi$ ($\Bar{\Psi}$) is (anti)chiral
and $\eta$, $\Bar{\eta}$, $Y$, $\Bar{Y}$ are unconstrained complex
superfields.  The superfields $Y$ and $\Bar{Y}$ act as complex
Lagrange multipliers for the nonminimal part of the CNM
constraints (\ref{defkinCNM}).  In fact, integrating out $Y$ and
$\Bar{Y}$, the action (\ref{CCNMdual}) reduces to (\ref{SCNM})
with $\F \equiv\Psi$ and $\S\equiv\eta$ which are now constrained
to satisfy the conditions (\ref{defkinCNM}).  On the other hand,
if we integrate out $\eta$ and $\Bar{\eta}$  using the equations of
motion $\eta=D^2\Bar{Y}$, $\Bar{\eta}=\Db^2Y$, and define the chiral
superfields $\O_-\equiv\Db^2Y$, $\Ob_-\equiv D^2\Bar{Y}$,
$\O_+\equiv \Psi$, $\Ob_+\equiv\Bar{\Psi}$, the action
(\ref{CCNMdual}) reduces to (\ref{SCC}) and describes a CC
hypermultiplet.

We note this kind of duality is {\em solely} due to the 4D superspace
structure which we use to define the multiplets and it should not be affected
by the spacetime dimension in which we are working.  As important it
should be noted this duality would likely {\em not} exist in a manifestly
6D $\cal N$ $=$ $1$ formulation.  So one or the other of these two distinct
but duality-related formulations may be preferred from this perspective.

Therefore, we expect in six dimensions more general duality patterns.  For
example, in \cite{CNM} the most general class of 4D CNM models with
constraints $\Db^2\S^a=Q^a(\F^b)$ was proposed, where $\S_a$ are
nonminimal superfields and $Q^a$ are holomorphic functions of a set
of chirals $\F^b$.   Using a simple generalization of the ordinary duality
transformations, it is possible to prove the four dimensional CNM models
described by the action
\beq
\int d^4x d^4\te\Big[\,\Fib_b\F^b-\Sb_a\S^a\,\Big]
\eeq
are dual to CC models with action
\beq
\int d^4x d^4\te\Big[\,\Ob_{+b}\O_+^b+\Ob_-^a\O_{-a}\,\Big]+\Bigg\{\int
d^4x d^2\te\,\O_{-a}Q^a(\O_+^b)+{\rm h.c.}\Bigg\}~~.
\eeq
It may be possible to extend this general duality to six dimensional models.
As the dimension of spacetime is varied, the class of functions $Q^a$ which are
used can also be subjected to different constraints.  For example, in four
dimensions the $Q$-functions may be used to introduce quartic
component-level self-couplings among the scalars.  It is unlikely that such
interactions are allowed in dimensions greater than four.   Still the
$Q$-functions are useful in other ways.  For the purposes of this work the
relevant choice is $Q=\pa\F$ which gives the 6D, $\CN=(1,0)$ hypermultiplet.

To conclude this section we note the descriptions of the $\CN=(1,0)$
hypermultiplet given above can be easily implemented in the case of a
hypermultiplet with opposite chirality.  To describe $\CN=(0,1)$ 
hypermultiplets it is only necessary to exchange $\pa$ with $-\pab$ in all the 
previous formulae. In fact, in this way the scalar parts of the actions 
(\ref{S0CC}, \ref{S0CNM}) do not change, whereas the spinors built from $\O_-$,
$\O_+$ and $\F$, $\S$ describe the free dynamics of a 6D $(0,1)$ Weyl spinor as
 in (\ref{free6d(0,1)}) (once again in the appendix), thus giving the correct 
physical content of a $\CN=(0,1)$ hypermultiplet.

\setcounter{equation}{0}
\section{Coupling Hypermultiplets to SYM}

~~~~On--shell formulations of higher dimensional supersymmetric Yang--Mills
theories, based on a formalism which keeps 4D supersymmetry manifest, have
been considered in literature for the 10D case \cite{MaSaSi}  and successively
for any dimension $D$, $5\leq D\leq 10$ \cite{ArkaniHamedTB}.  We will review
the results for the six dimensional case since we are eventually interested to
minimally couple matter described by the 6D, CNM hypermultiplets introduced
in the previous section.

Superspace covariant derivatives and field strengths can be constructed
\cite{MaSaSi} in terms of a real prepotential $V$ and a (anti)chiral superfield
$\O$ ($\Ob$). These are functions of the six bosonic coordinates and have an 
ordinary expansion in $(\q_\a, \bar{\q}_\ad)$, so realizing a representation of
4D supersymmetry. They belong to the adjoint representation of the gauge group,
 $V = V^i T_i$, $\O = \O^i T_i$ where $T_i$ are the group generators.  They are
subjected to the gauge transformations
\bea
e^{V'}&=&e^{i\Lb}e^{V}e^{-i\L}~,~
\nonumber \\
\O'=e^{i\L}\O e^{-i\L}+ie^{i\L}(\,\pa e^{-i\L}\,) ~&,&~
\Ob'=e^{i\Lb}\Ob e^{-i\Lb}+ie^{i\Lb}(\,\pab e^{-i\Lb}\,)~~,~
\label{SYMgs}
\eea
where $\L$ is a 4D chiral superfield depending also on the two extra 
coordinates $z,\bar{z}$.

In the chiral representation covariant derivatives are given by
\bea
&\Dcb_{\ad}=\Db_{\ad}~~~
,~~~\Dc_{\a}=D_{\a}-i\G_{\a}=e^{-V}D_{\a}e^{V}~~~,~~~
\Dc_{\a\ad}=-i\{\Dc_\a,\Dcb_\ad\}~~~,\non\\
&\Dc_{z}=\pa-i\G_{z}=\pa-i\O~~~,~~~
\Dcb_{\Bar{z}}=\pab-i\Bar{\G}_{\Bar{z}}=e^{-V}(\,\pab-i\Ob\,)e^{V}~~~,~~~
\label{CovDev6d}
\eea
and transform covariantly under the gauge transformations (\ref{SYMgs})
($\Dc_A \rightarrow e^{i\L} \Dc_A e^{-i\L}$, where 
$A=(\a ,\ad , \a\ad ,z ,\bar{z})$).
This geometric set--up is the generalization to six dimensions of the chiral
representation of the ordinary $\CN=1$ superspace gauge covariant derivatives
in four dimensions.  The superfields $\O$ and $\Ob$ play the role of 
connections associated with the two extra derivatives $\pa$, $\pab$.

The covariant derivatives satisfy the constraints
\bea
&&
F_{\a\b}=i\{\Dc_\a,\Dc_\b\}= 0 ~~~~~~, ~~~~~~
F_{\ad\bd}=i\{\Dcb_\ad,\Dcb_\bd\}=0~~,~
\nonumber \\
&& F_{\a\Bar{z}}= i[\Dc_\a,\Dcb_{\Bar{z}}]= 0 ~~~~~~~~, ~~~~~~
F_{\ad z}=i[\Dcb_\ad,\Dc_z]=0~~~~,~ \label{constrCovDev6d}
\eea
and give the set of non--trivial field strengths
\bea
W_{\a}&=&{i\over
2}[\Dcb^\ad,\{\Dc_\a,\Dcb_\ad\}]=i\Db^2(e^{-V}D_{\a}e^{V})
~~~,\\
W_{\ad}&=&{i\over 2}[\Dc^\a,\{\Dcb_\ad,\Dc_\a\}]=e^{-V}iD^2(e^{V}
\Db_{\ad}e^{-V})e^{V}~~~,\non\\
F_{\a z}&=&i[\Dc_\a,\Dc_z]=D_\a\O-i\pa(e^{-V}D_{\a}e^{V})
+[(e^{-V}D_{\a}e^{V}),\O]~~~,\non\\
F_{\ad\Bar{z}}&=&i[\Dcb_\ad,\Dcb_{\Bar{z}}]=i\Db_\ad (e^{-V}\pab e^{V})
+\Db_\ad(e^{-V}\Ob e^{V})~~~,\non\\
F_{z\Bar{z}}&=&i[\Dc_z,\Dcb_{\Bar{z}}]=i\pa(e^{-V}\pab e^{V})
+\pa(e^{-V}\Ob e^{V})-\pab\O+\non\\
&&~~~~~~~~~~~~~~~~~~~~~~+[\O,(e^{-V}\pab e^{V})]-i[\O,(e^{-V}\Ob e^{V})]~~~.
\non
\label{SYMfieldStregts6d}
\eea

The gauge invariant action in six dimensions is
($d^{10}Z\equiv d^6xd^4\te$, \textbf{$d^{8}Z\equiv d^6xd^2\te$})
\be \eqalign{ {~~~~~~~}
S_{SYM_{6}}[V,\Omega,\Bar{\Omega}] &=~ {1\over {2g^2}}\Tr \int d^{8}Z~W^\a
W_\a+{1 \over g^2}\Tr \int d^{10}Z\Bigg[~e^{-V}\Bar{\Omega}e^{V}\Omega \cr
&~~~~~~~~~~~~~~~~~~~~~~~~~~~+~
i(\pa e^{-V})\Bar{\Omega}e^{V}
~-~i e^{V}\Omega(\Bar{\pa} e^{-V}) \cr
&~~~~~~~~~~~~~~~~~~~~~~~~~~~
+~{1\over 2}(e^{-V}\Bar{\pa}
e^{V})(e^{-V}\pa e^{V})\cr
&~~~~~~~~~~~~~~~~~~~~~~~~~~~
+~(\Bar{\pa}V)\({{\sinh L_{V} - L_{V}}\over
{(L_{V})^2}}\)(\pa V)~\Bigg]~,~  }
\label{SYM6d}
\ee
where $g$ is the gauge coupling constant of dimension $-1$.
The equations of motion from its variation are
\bea
\{\Dc^\a,W_\a\}-{1\over
2}F_{z\Bar{z}}=0&,&\{\Dc^\a,F_{\a z}\}=0~~,
\eea

When reduced to components with the auxiliary fields integrated out, this
action describes the dynamics of a 6D, $\CN=(0,1)$ vector multiplet given
by a 6D vector field and a $(0,1)$ Weyl spinor \cite{MaSaSi,ArkaniHamedTB}.
The real superfield $V$ contains the 4D part of the 6D vector field and half
of the 6D Weyl spinor.  The connections $\O$ and $\Ob$ contain the rest of
the physical degrees of freedom \cite{ArkaniHamedTB,MaSaSi}.

By dimensional reduction, the previous action can be derived from
the ten dimensional $\CN=1$ supersymmetric action found in
\cite{MaSaSi}. Proceeding in this way what one finds is a 6D
$\CN=(1,1)$ SYM, where the vector multiplet is described by the
action (\ref{SYM6d}) and it is minimally coupled to a CC
hypermultiplet in the adjoint representation of the gauge group.
Setting to zero the hypermultiplet we are left with the action
for the $(0,1)$ vector multiplet.

As for the hypermultiplets, in order to describe a vector
multiplet with opposite chirality it is sufficient to exchange
$\pa\leftrightarrow-\pab$ in the previous formulation. In
particular, for the $\CN=(1,0)$ SYM we impose the constraints
\bea
&& F_{\a\b}=i\{\Dc_\a,\Dc_\b\}= 0 ~~~~~~, ~~~~~~
F_{\ad\bd}=i\{\Dcb_\ad,\Dcb_\bd\}=0~~,~
\nonumber \\
&& F_{\a z}= i[\Dc_\a,\Dc_{z}]= 0 ~~~~~~~~, ~~~~~~ F_{\ad
\Bar{z}}=i[\Dcb_\ad,\Dcb_{\Bar{z}} ]=0~~~~,~
\eea
whereas
$F_{\a\Bar{z}}$, $F_{\ad z}$ are non--trivial. In the chiral
representation for the 4D superspace covariant derivatives, these
constraints imply the definitions (\ref{CovDev6d}) are
modified as long as $(z, \Bar{z})$--derivatives are concerned,
according to
\beq
\Dc_{z}=\pa-i\G_{z}=e^{-V}(\pa-i\Ob)e^V~~~~~~,~~~~~~~
\Dcb_{\Bar{z}}=\pab-i\Bar{\G}_{\Bar{z}}=\pab-i\O ~~,
\eeq
Therefore, in the $\CN=(1,0)$ case the chiral connection is
$\Gb_{\Bar{z}}=\O$.

Now we study the minimal coupling of hypermultiplets to 6D SYM.
This can be accomplished by simply changing the definitions of the 
hypermultiplets to insure covariance under gauge transformations. Similarly to 
the 4D, $\CN=1$ case, this amounts to implement all the derivatives in the 
constraints and in the actions to be the gauge covariant derivatives in eq. 
(\ref{CovDev6d}).

The coupling of the 6D CC hypermultiplet to SYM has been given
in \cite{ArkaniHamedTB}. Following the same procedure it is easy
to couple the hypermultiplet when it is formulated in terms of CNM
superfields. Therefore, we treat the two cases together.

We consider $\CN=(1,0)$ covariantly CC and CNM hypermultiplets
belonging to a given representation of the gauge group and defined
by the following covariant constraints
\bea
&\Dcb_\ad
\Omega_{c\pm}=0~~~,~~~\Dc_\a\Bar{\Omega}_{c\pm}=0~~~,~~~
\label{covdefkinCC}\\
&\Dcb_\ad\F_c=0~~~,~~~\Dc_\a\Fib_c=0~~~,~~~\Dcb^2\S_c=\Dc_z\F_c~~~,~~~
\Dc^2\Sb_c=\Dcb_{\Bar{z}}\,\Fib_c~~~.
\label{covdefkinCNM}
\eea
The corresponding gauge invariant actions read
\beq
S_{CC}=\int
d^{10}Z \Big[\,\Bar{\Omega}_{c+}\Omega_{c+}+
\Bar{\Omega}_{c-}\Omega_{c-}\Big]+
\int d^8Z \Big[\,\Omega_{c+}\Dc_z\,\Omega_{c-}\Big]+
\int d^8\Bar{Z}\Big[\,\Bar{\Omega}_{c+}\Dcb_{\Bar{z}}\,\Bar{\Omega}_{c-}\Big]~,
\label{ScovCC}
\eeq
\beq
S_{CNM}=\int d^{10}Z\Big[~\Fib_c\F_c-\Sb_c\S_c~\Big]~.
\label{ScovCNM}
\eeq
As an interesting point, we note the constraints (\ref{constrCovDev6d})
on the covariant derivatives follow as consistency conditions for the
existence of $\CN=(1,0)$ covariant CNM hypermultiplets (\ref{covdefkinCNM}).
Therefore, $\CN=(1,0)$ CNM hypermultiplets can only be coupled to $\CN=
(0,1)$ SYM vector multiplets. This translates into a pure kinematic language,
the well--known fact in six dimensions, hypermultiplets with a given
chirality can only be coupled to vector multiplets of opposite chirality
\cite{ArkaniHamedTB,SUSYKHSTK6d}. In the case of the CC formulation of
hypermultiplets the same condition arises at the dynamical level, since the
action as written in eq. (\ref{ScovCC}) makes sense only if the derivative
$\Dc_z$ ($\Dcb_{\Bar{z}}$) does not spoil the chirality of the superfield
$\O_{c\pm}$ ($\Bar{\O}_{c\pm}$).

In the particular case of matter in the adjoint representation of the gauge
group we find it convenient to re--express the actions in terms of ordinary
(non--covariant) superfields. Therefore, given chiral superfields $\O_{\pm}$
and $\Phi$ satisfying $\Bar{D}_\ad \O_{\pm} = \Bar{D}_\ad \Phi =0$ and a
complex--linear multiplet with modified conditions
\bea
\Db^2\S=\pa\F-i\[\O,\F\]~~~&,&~~~D^2\Sb=\pab\,\Fib-i\[\Ob,\Fib\]~~~.
\label{defkinadCNM}
\eea
the previous actions can be re--written as
\be
~({\rm CC})~~~~~~~~~~~~~~~~
\eqalign{
&\Tr\int d^{10}Z \Big[ e^{-V}\Ob_{+}e^{V}\Omega_{+} +
e^{-V}\Bar{\Omega}_{-}e^{V}\Omega_{-}\Big]+  \cr
&
~~+ \Bigg\{\Tr\int d^8Z
\Big[\Omega_{+}\pa\,\Omega_{-}-i\,\O_+\[\O,\O_-\]\Big]+({\rm
h.c.})\Bigg\} ~~,}
\label{SadCC}
\ee
\be
({\rm CNM})~~~~~~~~~~~~~~\,\,
\Tr\int d^{10}Z \Big[~e^{-V}\Fib
e^{V}\F- e^{-V}\Sb e^{V}\S~\Big]~~,~\,\,\,\,\,\,\,~~~~~~~~~~~~~
\label{SadCNM}
\ee
respectively.  Using a gauge covariant generalization of the duality 
transformations described in section $2$, it is straightforward to prove the 
two theories are still dual.

Adding the action (\ref{SYM6d}) for the $\CN=(0,1)$ vector multiplet ($g=1$
for simplicity) to (\ref{SadCC}) we have the action describing 6D, $\CN=(1,1)$
SYM as obtained by dimensional reduction of the 10D $\CN=1$ SYM of
\cite{MaSaSi}.
Instead, if we add (\ref{SYM6d}) to (\ref{SadCNM}) we find an action for a
non--minimal representation of the 6D $\CN=(1,1)$ SYM.  Both the resulting 
theories, when dimensionally reduced to four dimensions, give the on--shell 
$\CN=4$ SYM, in the second case in a non--minimal representation.

\setcounter{equation}{0}
\section{Quantization of the 6D Multiplets}

~~~~In this section we perform the quantization of the 6D multiplets considered
in the previous sections, using a 4D, $\CN=1$ covariant procedure. We 
concentrate only on multiplets of a given 6D chirality, the cases with opposite
chirality being completely analogue.

\paragraph{CC Hypermultiplet.}

Given the $\CN= (1,0)$ hypermultiplet in the CC formulation, we
proceed to the quantization of its action (\ref{SCC}). In complete
analogy with the 4D, $\CN=1$ case \cite{SUPERSPACE}, we first solve
the (anti)chiral constraints (\ref{defkinCC}) by expressing
$\O_{\pm}$ ($\Bar{\O}_{\pm}$) in terms of two unconstrained
superfields
\bea
\O_\pm=\Db^2\psi_\pm&,&\Ob_\pm=D^2\Bar{\psi}_\pm~~.
\label{risvinCC}
\eea
The kinetic action (\ref{SCC}) becomes
\bea
S^{0}_{CC}&=&\int d^{10}Z~(~\Bar{\psi}_-~,~\psi_+~)
\left(\begin{array}{cc}
~D^2\Db^2 & -\pab D^2\vspace{1ex}\\
~\pa \Db^2&\Db^2D^2 \\
\end{array}\right)
\left( \begin{array}{c}
\psi_- \vspace{1ex}\\  \Bar{\psi}_+ \end{array} \right)~.
\label{SkinCC}
\eea
As a consequence of the invariance of the action under
$\d \psi_\pm=\Db^\ad\Bar{\c}_{\pm\ad}$
($\d \Bar{\psi}_\pm=D^\a\c_{\pm\a}$), the kinetic operator in
(\ref{SkinCC}) is not invertible.
We can fix these invariances using the well known gauge--fixing procedure for
the four dimensional massless scalar chiral superfield \cite{SUPERSPACE}.
This amounts to add gauge--fixing terms which complete the operators
$D^2\Db^2$ and $\Db^2D^2$ to $\Box_4$ (see \cite{SUPERSPACE} for details).
The kinetic action (\ref{SkinCC}) then reads
\bea
S^{0}_{CC}+S_{GF}&=&\int d^{10}Z~(~\Bar{\psi}_-~,~\psi_+~)
\left(\begin{array}{cc}
\Box_4&-\pab D^2\vspace{1ex}\\
\pa \Db^2&\Box_4 \end{array}\right)
\(\begin{array}{c}\psi_-\vspace{1ex}\\\Bar{\psi}_+\end{array}\)~.
\label{SkinCCfixed}
\eea
Moreover, as in the ordinary 4D case, at the end of the gauge--fixing procedure
the ghosts decouple from the physical superfields. The kinetic operator in
(\ref{SkinCCfixed}) is now invertible and from its inverse we find the
following propagators
\beq
\left(\begin{array}{cc}   <\psi_-\Bar{\psi}_->&<\psi_-\psi_+>\vspace{1ex}\\
<\Bar{\psi}_+\Bar{\psi}_->&<\Bar{\psi}_+\psi_+> \end{array}\right)
~=~
{1\over \Box_4}
\left(\begin{array}{cc}
~\({D^2\Db^2\over\Box_4}\frac{\pa\pab}{\Box_6}-1\)&
-{\pa D^2\over\Box_6}
\vspace{1ex}\\~{\pab\,\Db^2\over\Box_6}&
\({\Db^2D^2\over\Box_4}\frac{\pa\pab}{\Box_6}-1\)
\end{array}\right)
~.
\label{invkinCCfixed}
\eeq
Using the definitions (\ref{risvinCC}), we finally have the propagators for
the physical superfields
\bea
<\O_-\Ob_->~=\,-{\Db^2D^2\over\Box_6}\d^{10}(Z-Z')&,&
<\O_-\O_+>~=\,-{\pa \Db^2\over\Box_6}\d^{10}(Z-Z')~~~,\non\\\non\\
<\Ob_+\Ob_->~=~{\pab D^2\over\Box_6}\d^{10}(Z-Z')~~~~&,&
<\Ob_+\O_+>~=\,-{D^2\Db^2\over\Box_6}\d^{10}(Z-Z')~.~~~~~~
\label{MatterCC}
\eea

\paragraph{CNM Hypermultiplet.}

In order to perform the quantization of the CNM hypermultiplet we
start from the action (\ref{SCNM}) for $\Phi$ chiral and $\S$
satisfying the more general constraints
\beq
\Db^2\S=\pa\F+\F
P(\F,\O_a)~~~,~~~
D^2\Sb=\pab\,\Fib+\Fib\,\Bar{P}(\Fib,\Bar{\O}_{\Bar{a}})~,~
\label{defkin2CNM}
\eeq
where $\O_a$ ($\Bar{\O}_{\Bar{a}}$) are a
set of (anti)chiral superfields and $P$ ($\Bar{P}$) is a
(anti)holomorphic function, analytic in the superfields, with an
expansion which starts from the linear order. The constraints
(\ref{defkinCNM}) and (\ref{defkinadCNM}) are particular cases of
(\ref{defkin2CNM}), with $P=0$ and $\F P(\F,\O)= i[\F,\O]$,
respectively.

In order to quantize the action (\ref{SCNM}), we follow closely the
procedure used in \cite{CNM0404222} for the 4D case. First of all
we solve the kinematical constraints (\ref{defkin2CNM}) which
define $\F$ and $\S$. The most general solution is given in terms
of unconstrained superfields as
\bea
\F\equiv\Db^2\c~~~&,&~~~\S=\Db^\ad\sba_\ad+\pa\c+\c P(\F,\F_a)~~,
\label{rvCNM1}\\
\Fib\equiv D^2\Bar{\c}~~~&,&~~~\Sb=D^\a\s_\a+
\pab\,\Bar{\c}+\Bar{\c} \Bar{P}(\Fib,\Fib_{\Bar{a}})~~.
\label{rvCNM2}
\eea
The action (\ref{SCNM}) then reads
\beq
S_{CNM}=\int d^{10}Z\Big[\big(D^2\Bar{\c}\big)\big(\Db^2\c\big)-
\Big(D^\a\s_\a+\pab\,\Bar{\c}+\Bar{\c}\Bar{P}\Big)
\Big(\Db^\ad\sba_\ad+\pa\c+\c P\Big)\Big]~,~
\label{Siq}
\eeq
whose quadratic part is
\beq
S^0_{CNM} =\int
d^{10}Z~(~\Bar{\c}~,~\s_\a~) \left(\begin{array}{cc}
(D^2\Db^2+\pab\pa)&\pab\,\Db^\ad
\vspace{1ex}\\
-\pa D^\a&-D^\a\Db^\ad \end{array}\right)
\(\begin{array}{c}\c\vspace{1ex}\\
\sba_\ad\end{array}\)~.
\label{S0iq}
\eeq
The expressions
(\ref{rvCNM1}, \ref{rvCNM2}) and the action (\ref{Siq}) are
invariant under the following two sets of tranformations
\bea
\d\c=\Db^\ad\Bar{\c}_\ad~&,&~\d\sba_\ad=-\pa\,\Bar{\c}_\ad-\Bar{\c}_\ad
P(\F,\F_a)~,
\label{gaugechir}
\eea
and
\bea
\d\c&=&0~,\non\\
\delta \sigma^\alpha&=& D_\b \sigma^{(\b\alpha)}~,\nonumber\\
\delta \sigma^{(\b\alpha)}  &=& D_\gamma \sigma^{(\gamma\b\alpha)}~,
\nonumber\\
\delta \sigma^{(\gamma\b\alpha)}  &=& D_\delta
\sigma^{(\delta\gamma\b\alpha)}~,
\nonumber\\
~&\vdots&~\non\\
\d\s^{(\a_n\a_{n-1}\cdots\a_1)}&=&D_{\a_{n+1}}\s^{(\a_{n+1}\a_n\a_{n-1}
\cdots\a_1)}~,\non\\
~&\vdots&
\label{gaugetr}
\eea
Therefore, the kinetic operator in (\ref{S0iq}) is not invertible.

Since these invariances are due to the four--dimensional superspace structure
of the covariant derivatives, we can apply the gauge--fixing procedure of
\cite{CNM0404222} forgetting we are working in six dimensions.

The gauge--fixing procedure runs in two steps.
First, we consider the transformations (\ref{gaugechir}). As for the CC
hypermultiplet, to fix this invariance we use the standard gauge--fixing
procedure
of the four dimensional massless scalar chiral superfield \cite{SUPERSPACE}
to bring the kinetic operator $(D^2\Db^2+\pab\pa)$ in (\ref{S0iq}) to
$(\Box_4+\pab\pa)=\Box_6$.

As a second step we consider the transformations (\ref{gaugetr}).
Since $\c$ and $\Bar{\c}$ do not transform, we can use the
gauge--fixing procedure of \cite{QuantizedNM} for the ordinary
four dimensional complex--linear superfield and modify only the
$\s_\a$, $\Bar{\s}_\ad$ part of the kinetic action.
Precisely, the gauge--fixing is performed by introducing an infinite
tower of ghosts according to a non--trivial
superspace version of the Batalin--Vilkovinsky formalism.
As proved in \cite{QuantizedNM}, if the gauge--fixing functions are chosen
to be independent of the background physical fields, the tower of
ghosts can be completely decoupled by a finite number of ghost fields
redefinitions, and we are left
with an invertible kinetic term for $\s_\a$, $\sba_\ad$.

The same
procedure can be applied without modifications to our case. The
result is the conversion of the operator $D^\a\Db^\ad$ into the
invertible operator
\beq
W^{\a\ad}=\Bigg[~D^\a\Db^\ad+{k^2\over 2}\Db^\ad D^\a -{k^2\over
2}\({1+{k'_1}^2\over
1-{k'_1}^2}\)i\pa^{\a\ad}{\Db^2D^2\over\Box_4} +{k^2\over
2}i\pa^{\a\ad}{D_\b\Db^2D^\b\over\Box_4} ~\Bigg]~,
\label{Waad}
\eeq
where $k$ and $k_1'$ are two gauge--fixing parameters.

At the end of the procedure the gauge--fixed action reads
\beq
S_{CNM}+S^{tot}_{GF}=\int d^{10}Z~ (~\Bar{\c}~,~\s_\a~)
\left(\begin{array}{cc} ~\Box_6~&~\pab\,\Db^\ad
\vspace{1ex}\\~-\pa D^\a~&~-W^{\a\ad}~
\end{array}\right)
\(\begin{array}{c}\c \vspace{1ex}\\ \sba_\ad\end{array}\)~.
\label{StotCNMm}
\eeq
Inverting the kinetic operator (\ref{StotCNMm}) we find
\beq
\left(\begin{array}{cc} <\c\Bar{\c}>&<\c\s_\a>\\
<\sba_\ad\Bar{\c}>&<\sba_\ad\s_\a>
\end{array}\right)
= \left(\begin{array}{cc}
\frac{1}{\Box_4}\({\pab\pa\over\Box_6}{D^2\Db^2\over\Box_4}-1\)\,&\,
{\pab\over\Box_4}\(\frac{1}{2}{\Db^2D_\a\over\Box_4}-{D_\a\Db^2\over\Box_4}\)
\\\,&\,\\{\pa\over\Box_4}\(\frac{1}{2}{\Db_\ad D^2\over\Box_4}-
{D^2\Db_\ad\over\Box_4}\)\,&\,
W^{-1}_{\a\ad}+{\pab\pa\over\Box_4}W^{-1}_{\b\ad}D^\b\Db^\bd W^{-1}_{\a\bd}
\end{array}\right)~,
\label{propR}
\eeq
where
\be
\eqalign{
W^{-1}_{\a\ad}~=&-{i\pa_{\a\ad}\over\Box_4}~+~
{3(kk'_1)^2+4-2{k'_1}^2\over
4(kk'_1)^2}i\pa_{\a\ad}{\Db^2D^2\over\Box_4^2} ~+~
{3k^2-2\over4k^2}i\pa_{\a\ad}{D_\b\Db^2D^\b\over\Box_4^2}~+ \cr
&+~ {2-k^2\over 4k^2}i\pa_{\a\bd}i\pa_{\b\ad}{\Db^\bd
D^\b\over\Box_4^2}~~~,}
\ee
is the inverse of $W^{\a\ad}$.

In the particular case of CNM multiplet in (\ref{defkinCNM}),
$P=\Bar{P}\equiv0$, we easily infer the propagators of the
physical superfields
\bea
<\F\Fib>\,=\,-{\Db^2D^2\over\Box_6}\d^{10}(Z-Z')&,&\,
<\S\Fib>\,=\,-{\pa D^2\over\Box_6}\d^{10}(Z-Z')~~~~~~~~~~,~~~\non\\\non\\
<\F\Sb>\,=\,{\pab\,\Db^2\over\Box_6}\d^{10}(Z-Z')~~~~&,&\,
<\S\Sb>\,=\,\Bigg[1-{D^2\Db^2\over\Box_6}\Bigg]\d^{10}(Z-Z')
~~.~~~~~~\label{MatterCNM}
\eea

\paragraph{Vector multiplet.}

The quantization of the $\CN=(0,1)$ vector multiplet can be
performed by following closely the procedure described in
\cite{MaSaSi} for the 10D case. For simplicity we set $g=1$
in (\ref{SYM6d}).

The quadratic part of the action (\ref{SYM6d}) is
\beq
S^{(2)}_{SYM_{6}}[V,\Omega,\Bar{\Omega}]= \Tr \int
d^{10}Z\Big[-{1\over 2}VD_\a\Db^2D^\a V+ \Bar{\Omega}\Omega-i(\pa
V)\Bar{\Omega}+i \Omega(\Bar{\pa} V)+ {1\over 2}(\Bar{\pa} V)(\pa
V)\Big]~.~
\label{quadSYM6d}
\eeq
invariant under the gauge
transformations (\ref{SYMgs}). To fix this invariance we choose a
Feynman--type gauge--fixing term \cite{MaSaSi} suitably adapted to
the six dimensional case
\beq
S_{GF}=-\Tr\int
d^{10}Z\Bigg(\Db^2V+i\,{\Db^2\over \Box_4}\,\pa\,\Ob\Bigg)
\Bigg(D^2V-i\,{D^2\over \Box_4}\,\pab \, \O\Bigg)~.
\label{SYMgfterm}
\eeq
The corresponding Faddev--Popov ghosts
action is
\bea
S_{FP}&=&-\Tr\int d^{10}Z\Bigg[
(c'+\Bar{c}')L_{\frac{V}{2}}\Big((c+\Bar{c}) +\coth
L_{\frac{V}{2}} (c-\Bar{c})\Big)-
c'\,{\pa\pab\over \Box_4}\,\Bar{c}\non\\
&&~~~~~~~~~ -i(\pa c'){1\over
\Box_4}[\Ob,\Bar{c}]+\Bar{c}'\,{\pa\pab\over \Box_4}\,c +i(\pab\,
\Bar{c}'){1\over \Box_4}[\O,c]\Bigg]~.
\eea
The advantage of using
the gauge--fixing term (\ref{SYMgfterm}) is in the
quadratic part of the action the superfield $V$ decouples from $\O$ and $\Ob$
\beq
S^{(2)}=\Tr\int d^{10}Z\Bigg[-{1\over 2}V\Box_6V+
\Ob\,{\Box_6\over \Box_4}\,\O +{\rm Ghosts}\Bigg]~.
\label{S2totSYM}
\eeq
and the propagators in the chosen gauge are
\bea
<VV>={1\over \Box_6}\d^{10}(Z-Z')~~\,&,&~<\Ob\O>=-{D^2\Db^2\over
\Box_6}\d^{10}(Z-Z')~~~.
\label{propSYM6d}
\eea

We conclude this section with few comments. Given the particular
approach we have used to study six dimensional superfields,
the quantization turns out to be not much affected
by working in six dimensions and the results are very similar
to the four--dimensional case. In particular,
a formal equivalence between 4D CC/CNM massive propagators and
6D CC/CNM massless ones can be established
by identifying the 4D complex mass with the extra dimensions
derivative operators, $m\leftrightarrow\pa$,
$\Bar{m}\leftrightarrow -\pab$. In the case of the vector
multiplet the corrispondence would work with a 4D massive
vector multiplet written in a superspace Stueckelberg
formalism \cite{BuchbinderKuzenko}.
The previous correspondence could be very useful
when studying quantum properties of six dimensional theories
in a 4D $\CN=1$ formalism.

\setcounter{equation}{0}
\section{The 6D Projective Superspace Perspective}

~~~~In four dimensions, the complex--linear superfield plays an
important role in the definition of $\CN=2$ multiplets in the
context of projective superspace
\cite{ProjectiveSuperspaceGR,ProjectiveSuperspace,
ProjectiveSuperspaceK,ProjectiveSuperspaceGK}.
In fact, the on--shell $\CN=1$ superspace description of the
$\CN=2$ (ant)artic projective superfield is given in terms of a 4D
CNM multiplet
\cite{ProjectiveSuperspaceGR,ProjectiveSuperspace,
ProjectiveSuperspaceK,ProjectiveSuperspaceGK}.
Having constructed 6D hypermultiplets using a CNM
multiplet, it is then natural to ask if a projective superspace
formulation of 6D supersymmetry exits and if one can define
there 6D superfields whose on--shell version is given by the
multiplets previously introduced.

Since we have defined 6D multiplets always keeping manifest only
the 4D superspace structures, we try to formulate the 6D
projective superspace using the same approach\footnote{Recently,
in \cite{KuzenkoLinch} a similar extension has been found for the
five dimensional case. Note also that \newline
$~~~~~$ a previous formulation of 6D projective 
superspace was provided in \cite{GrunLind} to describe the $O(2)$ 
\newline $~~~~~$
tensor multiplet in six dimensions.}. 
In this way we take advantage of the fact
that the 4D projective superspace manifestly preserves many
structures of the ordinary $\CN=1$ superspace.

First of all we remind our readers the algebra of the $\CN=(1,0)$
supercovariant derivatives is
\beq
\{D^{a\tilde{\a}},D^{b\tilde{\b}}\}=
\e^{ab}\G^{\m\tilde{\a}\tilde{\b}}i\pa_\m
~~~,\label{6dSUSYsupcovdev}
\eeq
where
$\G^{\m\tilde{\a}\tilde{\b}}$ have been defined in the appendix,
$\e^{ab}$ is the invariant tensor of the SU($2$) automorphism
group of the $\CN=(1,0)$ algebra and the derivatives
$D^{a\tilde{\a}}$ are $(1,0)$ Weyl spinors satisfying a
SU($2$)--Majorana condition \cite{SUSYKHSTK6d}. For our purposes
the algebra (\ref{6dSUSYsupcovdev}) can be written using the 4D
spinor notation as
\bea
&\{D_{a\a},D_{b\b}\}=\e_{ab}C_{\a\b}\pab~~~,~~~
\{\Db^a_{\ad},\Db^b_{\bd}\}=\e^{ab}C_{\ad\bd}\pa~~~,\non\\
&\{D_{a\a},\Db^b_{\bd}\}=\d_a^bi\pa_{\a\bd}~~~.
\label{6dSUSYsupcovdev4Df}
\eea
The interesting point is this
algebra has the same structure of the 4D $\CN=2$ algebra with a
complex central charge given by ($\e_{ab}\pab$). Therefore, we can
generalize to six dimensions the construction of the projective
superspace for the case of an underlying 4D $\CN=2$ SUSY with
central charge, as given in \cite{ProjectiveSuperspaceGR}.

We parametrize the projective superspace with a complex coordinate
$\z$ and we define the projective supercovariant derivatives as
\bea
\Dc_\a(\z)=\z D_{1\a}-D_{2\a}&,& \Dcb_\ad(\z)=\Db^1_\ad+\z
\Db^2_\ad~~~,
\label{projcovdev}
\eea
with the orthogonal set of
supercovariant derivatives given by
\bea
\DD_\a(\z)=D_{1\a}+{1\over \z}D_{2\a}&,&
\DDb_\ad(\z)=\Db^2_\ad-{1\over \z}\Db^1_\ad~~~.
\label{ortprojcovdev}
\eea
The projective supercovariant
derivatives algebra is
\bea
&\{\Dc_\a(\z),\Dc_\b(\z)\}=\{\Dc_\a(\z),\Dcb_\ad(\z)\}=
\{\Dcb_\ad(\z),\Dcb_\bd(\z)\}=0~~~,\non\\
&\{\DD_\a(\z),\DD_\b(\z)\}=\{\DD_\a(\z),\DDb_\ad(\z)\}=
\{\DDb_\ad(\z),\DDb_\bd(\z)\}=0~~~,
\non\\
&\{\Dc_\a(\z),\DD_\b(\z)\}=2C_{\a\b}\pab~~~,~~~
\{\Dcb_\ad(\z),\DDb_\bd(\z)\}=2C_{\ad\bd}\pa~~~,\non\\
&\{\Dc_\a(\z),\DDb_\ad(\z)\}=-\{\Dcb_\ad(\z),\DD_\a(\z)\}=-2i\pa_{\a\ad}~~~.
\label{projScov}
\eea

Following
\cite{ProjectiveSuperspaceGR,ProjectiveSuperspace,ProjectiveSuperspaceK,
ProjectiveSuperspaceGK}, superfields living on the projective
superspace are constrained by
\bea
\Dc_\a(\z)\Xi=0=\Dcb_\ad(\z)\Xi~~~\Longrightarrow~~~ D_{2\a}\Xi=\z
D_{1\a}\Xi~,~\Db^1_\ad\Xi=-\z\Db^2_{\ad}\Xi~~.
\label{constrProjSupf}
\eea
Now, the projective superfield $\Xi$
is a function of the six bosonic coordinates, of the
grassmannian $(\q^{a\a}, \Bar{\q}_{a}^{\ad})$ and it is chosen to
be holomorphic in $\z$ on ${\mathbb C}^*$. It can be expanded as
\beq
\Xi(x_i,\q^{a\a},
\Bar{\q}^\ad_a,\z)=\sum_{n=-\infty}^{+\infty}\Xi_n(x_i,\q^{a\a},
\Bar{\q}^\ad_a)\,\z^n~~~,
\label{ProjSupSerie}
\eeq
where $\Xi_n$ are $\CN=2$ superfields satisfying
\bea
D_{2\a}\Xi_{n+1}=D_{1\a}\Xi_{n}~~~,~~~
\Db^2_{\ad}\Xi_{n}=-\Db^1_{\ad}\Xi_{n+1}~~,
\label{compConstrProjSupf}
\eea
as follows from eqs. (\ref{constrProjSupf}).
The above constraints fix the dependence
of the $\Xi_n$ on half of the Grassmannian coordinates
($\te^{a\a},{\Bar \q}_a^\ad$) of the superspace. Therefore, $\Xi_n$ can
be considered as superfields which effectively live on a $\CN=1$
superspace with for example $\te^\a=\te^{1\a}$, $\teb^\ad={\Bar
\q}_1^\ad$
\cite{ProjectiveSuperspaceGR,ProjectiveSuperspace,ProjectiveSuperspaceK,
ProjectiveSuperspaceGK}.

In projective superspace the natural conjugation operation
combines complex conjugation with the antipodal map on the Riemann
sphere ($\z\to-1/\z$) and acts on projective superfields as
\beq
\Xis=\sum_{n=-\infty}^{+\infty}\Xis_n\,\z^n=
\sum_{n=-\infty}^{+\infty}(-1)^n\Xib_{-n}\,\z^n~~~.
\label{smileconjugation}
\eeq

Similarly to the 4D case
\cite{ProjectiveSuperspaceGR,ProjectiveSuperspace,ProjectiveSuperspaceK,
ProjectiveSuperspaceGK}, six dimensional $\CN=(1,0)$ SUSY
invariant actions have then the general form
\beq
\int d^6x\ggl{1\over 2\pi i}\oint_C{d\z\over\z}
D^2\Db^2{\cal{L}}(\Xi,\Xis,\z)\Big|\ggr~~,
\eeq
where ${\cal{L}}(\Xi,\Xis,\z)$ is real under the
$\smile$-conjugation (\ref{smileconjugation}) and $C$ is a contour
around the origin of the complex $\z$--plane.

We have constructed a projective superspace which seems to have
the right properties to generalize the $\CN=1$ formalism used in
the previous sections to study 6D supersymmetric models. The
non--trivial property of this formulation is the linear
realization of the 6D, $\CN=(1,0)$ supersymmetry. Now we construct
6D projective superspace multiplets which have the same physical
content as the CNM multiplet and the vector multiplet previously
considered.

\paragraph{Polar Hypermultiplet.}

We study the polar hypermultiplet described by (ant)artic
superfields. In 4D it is the natural generalization of the CNM
multiplet. We define the artic and antartic projective superfields
as
\bea
\Y=\sum_{n=0}^{+\infty}\Y_n\,\z^n&,&
\Ys=\sum_{n=0}^{+\infty}(-1)^n\Yb_n\,{1\over \z^n}~~.
\label{(ant)artic}
\eea
Due to the truncation of the series, the
$\CN=1$ constraints on the component superfields $\Y_n$ are
\bea
\Db_\ad\Y_0=0&,&\Db^2\Y_1=\pa\,\Y_0~~,\non\\
D_\a\Yb_0=0&,&D^2\Yb_1=\pab\,\Yb_0~~,
\label{polarConstr}
\eea
with $\Y_n$ ($\Yb_n$) $n>1$ unconstrained $\CN=1$ superfields. The
natural action for a free polar multiplet is then
\beq
\int
d^6xd^4\te\ggl{1\over 2\pi i}\oint_C{d\z\over \z}\,\Ys\Y\ggr= \int
d^6xd^4\te\ggl\sum_{n=0}^{+\infty}(-1)^n\Yb_n\Y_n\ggr~~~.
\label{polarAction}
\eeq
The polar multiplet describes a
generalization of the 6D $\CN=(1,0)$ CNM hypermultiplet introduced
in section $2$, once we identify $\F\equiv\Y_0$
($\Fib\equiv\Yb_0$) and $\S\equiv\Y_1$ ($\Sb\equiv\Yb_1$) in
complete analogy with the four dimensional case. The infinite set
of auxiliary superfields in the polar multiplet extend the CNM
hypermultiplet to a multiplet which transforms linearly under 6D
$\CN=(1,0)$ supersymmetry.

\paragraph{Tropical Multiplet.}

Another interesting multiplet to consider in the framework of
projective superspace is the real tropical multiplet
\cite{ProjectiveSuperspaceGR,ProjectiveSuperspace,ProjectiveSuperspaceK}.
This is defined in terms of a projective superfield
$V(\z,\Bar{\z})$ which is analytic away from the poles of the
Riemann sphere and real under the $\smile$-conjugation. Therefore,
its expansion reads
\bea
V=\sum_{-\infty}^{+\infty}V_n\,\z^n&,&V_{-n}=(-1)^n\Bar{V}_n~~.
\label{tropicalV}
\eea
In four dimensions, the real tropical
multiplet is the prepotential of $\CN=2$ SYM. In the abelian case
the explicit expression for the action is known\footnote{To our
knowledge the tropical multiplet action is not explicitly known in the 
nonabelian.\newline $~~\,~~~$ In any case, it 
is important to note in
\cite{ProjectiveSuperspaceK} a prescription was given to extract
the entire\newline $~~\,~~~$ nonabelian
SYM action from the one written in
harmonic superspace.}
\cite{ProjectiveSuperspaceGR,ProjectiveSuperspace,ProjectiveSuperspaceK}.

Inspired by the 4D case, we consider the following action for a
real tropical projective superfield
\beq
S_{PSYM_6}=-{1\over
2}\int d^6xd^8\te \oint {d\z_1\over 2\pi i}{d\z_2\over 2\pi i}
{V(\z_1)V(\z_2)\over (\z_1-\z_2)^2}~~~.
\label{PSYM6}
\eeq
This is
formally equal to the action for a free real tropical projective superfield
in 4D \cite{ProjectiveSuperspaceGR,ProjectiveSuperspace,ProjectiveSuperspaceK}
with the only difference now that the fields live in a 6D superspace.  Now,
using the general constraints (\ref{constrProjSupf}) together with the 
identities
\bea
Q^2\Xi=\z^2D^2\Xi+\z\pab\,\Xi&,&
\Qb^2\Xi={1\over \z^2}\Db^2\Xi-{1\over \z}\pa\,\Xi~~,
\label{usfform}
\eea
where we have defined $D_\a=D_{1\a}~$,
$\Db_\ad=\Db^1_{\ad}~$, $Q_\a=D_{2\a}~$, $\Qb_\ad=\Db_\ad^2~$,
from (\ref{PSYM6}) we find
\bea
S_{PSYM_6} = -{1\over 2}\int d^6x
D^2\Db^2 \oint {d\z_1\over 2\pi i}{d\z_2\over 2\pi i} {1\over
(\z_1-\z_2)^2}\Big(Q^2\Qb^2V(\z_1)V(\z_2)\Big)\Big|=
~~~~~~~~~~~~~~~~~~~~~~~~~~~~~~~~~ \non
\eea
\beq
=\int d^{10}Z\Big[\,{1\over 2}V_0 D^\a\Db^2 D_\a
V_0-V_{-1}D^2\Db^2V_{1} -V_0\,\pa
D^2V_{-1}-V_1\,\pab\,\Db^2V_0-{1\over 2}V_0\,\pa\pab\,
V_0\,\Big]~.~~
\label{2PSYM6}
\eeq
Under the identification
$V\equiv V_0$, $\O\equiv i\Db^2V_1$, $\Ob\equiv
iD^2V_{-1}=-iD^2\Bar{V}_1$, the action (\ref{2PSYM6}) coincides
with (\ref{SYM6d}) for the abelian case. Then, as expected, the
action (\ref{PSYM6}) describes the dynamics of an abelian
$\CN=(0,1)$ vector multiplet in six dimensions.

The minimal coupling of general SYM gauge multiplets to
hypermultiplets in 6D projective superspace can be realized as in
the 4D case
\cite{ProjectiveSuperspaceGR,ProjectiveSuperspace,ProjectiveSuperspaceK}.
For example, the action for a polar multiplet in the fundamental
representation of the gauge group coupled to a tropical multiplet
is given by
\beq
\int d^6xd^4\te\ggl{1\over 2\pi
i}\oint_C{d\z\over \z}\,\Ys e^V\Y\ggr~~~.
\eeq
The action is
invariant under the gauge transformations
\beq
\Y'=e^{i\L}\Y~~~,~~~\Ys'=\Ys e^{-i{\breve{\L}}}~~~,~~~
(e^V)'=e^{i{\breve{\L}}}e^V e^{-i\L}~~,
\eeq
where the gauge
parameters $\L$ and $\breve{\L}$ are respectively artic and
antartic projective superfields.  The 6D projective superspace
description of multiplets with opposite chirality goes along the
same lines of the previous construction with $\pa$ exchanged with
$-\pab$.

We now address the issue of covariant quantization in
6D projective superspace.
To this end we expect the equivalence between the $\CN=(1,0)$
algebra written in the 4D formalism and the algebra of 4D $\CN=2$
supersymmetry with complex central charge should be still relevant in order
to extend the 4D results to six dimensions.

As an example, we concentrate on the polar multiplet. In
\cite{ProjectiveSuperspaceGR} the quantization of the 4D polar
multiplet was performed in the case of an underlying $\CN=2$
supersymmetry with complex central charge $m$. Exploiting the
formal identification\footnote{Note that this is the
identification which was pointed out at the end of section 4
between \newline $~~\,~~~$ the 6D $\pa$--derivative and the 4D
complex mass $m$. The only difference is in the
present  \newline
$~~\,~~~$ context the mass plays the role of the central charge.}
$m\leftrightarrow\pa$ we can easily argue the covariant
propagator for the 6D (ant)artic superfield. It is in fact
sufficient to take the result of \cite{ProjectiveSuperspaceGR} and
rewrite it in six dimensions with the correct insertions of $\pa$
and $\pab$. The propagator for the 6D $\CN=(1,0)$ polar multiplet
we expect then to be
\beq
<\Ys(Z,\z)\Y(Z',\z')>={1\over
\z'(\z-\z')^3} {\Dc^2(\z)\Dcb^2(\z)\Dc^2(\z')\Dcb^2(\z')\over
\Box_6}\,\d^8(\te-\te') \d^6(x-x')~,
\eeq
where
$Z=(x_i,\te^{a\a},\teb^\ad_a)$ are the coordinates of the 6D
superspace and the $\Dc_\a(\z)$, $\Dcb_\ad(\z)$ have been defined
in (\ref{projScov}).

For the 4D tropical multiplet in $\CN=2$ projective superspace
with central charge the quantization has not been performed
yet. Therefore, we cannot exploit the 4D results to easily infer the
form of its propagator in six dimensions.  In any case, the covariant
quantization in 6D projective superspace along the lines of
\cite{ProjectiveSuperspaceGR} has yet to be rigorously
developed.

So far we have restricted our attention to the polar and
tropical multiplets as projective superspace extensions of the 6D
CNM hypermultiplet and vector multiplet, respectively.
In standard 4D projective
superspace a larger class of multiplets has been studied and
classified. The classification is based on the analiticity
properties of the projective superfields in the $\z$--plane.
Projective superfields with a finite series expansion in $\z$
produce 4D complex $O(p)$ and real $O(2n)$ tensor multiplets. In
the limits $p\to\infty$, $n\to\infty$ these give the polar and
tropical multiplets, respectively
\cite{ProjectiveSuperspaceGR,ProjectiveSuperspace,ProjectiveSuperspaceK}.
Since our 6D projective superspace is essentially defined in the same way
as the 4D one (in particular for what concerns the $\z$--plane) it is clear the
same class of tensor multiplets can be easily constructed also in the 6D case.

\setcounter{equation}{0}
\section{Conclusions and Outlooks}

In these notes, using a formalism which keeps manifest the 4D
$\CN=1$ supersymmetry, we have introduced a new formulation of 6D
$\CN=1$ hypermultiplet in terms of chiral--nonominimal (CNM)
superfields. The CNM formulation is dual to the
chiral--chiral (CC) description already present in the literature
\cite{ArkaniHamedTB}. We
have coupled the CC and CNM hypermultiplets to 6D SYM,
covariantly with respect to the geometry of the 4D,
$\CN=1$ superspace. We have studied in detail the
superfield quantization of all the previous multiplets.
Furthermore, we have developed a 6D projective superspace
formalism in which the 6D CNM and vector multiplets naturally
emerge. We have also discussed the covariant quantization
of the (ant)artic projective superfields.

Armed with these results it would be interesting to investigate quantum
properties of 6D supersymmetric models. The advantage of using
a 4D, $\CN=1$ superfield formulation is in the possibility to
compare diagrams which arise in the 6D case to the 4D analogues
largely studied in the literature. This powerful
technique has been already used in \cite{MaSaSi} to study one--loop
properties of the ten dimensional $\CN=1$ SYM in correspondence to
the four dimensional $\CN=4$ SYM. Through the introduction of a 6D
projective superspace we have also
established a formalism which could be even more efficient for
exploring quantum properties of vector and hyper--multiplets in 6D.
For example it might be possible to exploit these formalism to
extend the study of 6D gauge anomalies \' a la previous work
\cite{anomSUSY}.

An interesting issue which might be worth studying in detail is
the relation between 6D projective and harmonic superspaces
\cite{HarmonicSuperspace}, along
the lines of \cite{ProjectiveSuperspaceK} in the 4D case. 
From the harmonic superspace perspective, 6D is interesting being
the highest dimension in which the powerful standard harmonic approach can be
used.
We expect the
relation between 6D harmonic \cite{harmonic6d} and projective
superspaces to have no relevant differences from the
4D case. 
The polar multiplet will be the projective superspace
version of the $q^+$ hypermultiplet and the tropical multiplet
will be related to the analytic harmonic gauge prepotential $V^{++}$.
Our expectation is also supported by the results recently obtained
in \cite{KuzenkoLinch} for the 5D case. In six dimensions
the only difference would be the fact, using the 4D spinor notation,
the SUSY algebra turns out to have a complex central charge. However, this
should not affect the structures which constrain the harmonics on one side
and the $\z$--complex--plane on the other one. Trying to understand the precise
formulation of nonabelian SYM in 6D projective superspace from the
harmonic one might be a useful indirect approach.\\
Recently in \cite{IvSmZu} using 6D harmonic superspace there was given the
action of a renormalizable higher derivatives 6D SYM theory. It
would be interesting to find the analogue of this theory written in 4D $\CN=1$
superfields formalism and projective superspace to study quantum properties of 
this model using our formalism.

Having a complete understading of the 6D harmonic superspace would be
very useful for addressing many issues. An interesting question to
investigate in this context would be how the ``harmonic
anomalies'' which arise in 4D quantum theories manifest
themselves in a 6D setting.
Furthermore, six dimensional harmonic superspace
might be the most efficient approach to analyze 6D $\CN=1$
nonlinear sigma--models in a completely covariant way.

The topic of six dimensional nonlinear sigma--models is an intriguing one
which has been not very well investigated to our knowledge. Since the
construction of 6D hypermultiplets and SYM is efficiently
developed using 4D $\CN=1$ superfields as ingredients, it is
natural to ask how to build supersymmetric sigma--models in this
formalism and what are their geometric properties
\cite{SigmaModels6D}.

In this respect the 6D projective superspace
formulation, rather than the harmonic one, should be the natural
starting point. In fact, once the reduction of the projective
superfields to their component superfields has been performed, we
obtain an action which is written in terms of 4D CNM $\CN=1$
superfields\footnote{A first example of how things should work can
be found in \cite{KuzenkoLinch} where the construction of 4D\newline $~~\,~~~$
CNM sigma-models \cite{ProjectiveSuperspaceGK} has been generalized to
five dimensions.}. The six--dimensional Lorentz invariance of this
action is not manifest. 
Therefore, one of the main
questions we need answer is how Lorentz invariance gets
restored once the model is reduced to the physical field
components. This is a good
starting point to attempt a formulation of supersymmetric CC sigma--models and 
CNM sigma--models which generalize the ones coming from projective superspace. 
6D Lorentz invariance imposes non--trivial geometrical constraints
on the sigma--model functions which describe the target space
manifold and brings to hyper-K\"ahler geometries. To this regard the CNM--CC 
duality is a really interesting issue. 

${~~~}$ \newline
${~~~~~}$``{\it {If you are out to describe the truth, leave
elegance to the tailor.}}'' \newline $~~~~~~~$ -- Albert Einstein

\section*{Acknowledgements}

\noindent
G.T.-M. thanks the Department of Physics of University of Maryland for the
kind and warm hospitality during the final stage of this work.

\newpage
\appendix

\setcounter{equation}{0}
\section{6D Weyl spinors}
\label{Notations}

In this section we introduce our notations and conventions for $6$D spinors.

In six dimensions, $(1,0)$ and $(0,1)$ Weyl spinors belong to the fundamental
representation of SU*($4$) and to the transpose representation, respectively.
These representations can be decomposed into 4D spinor representations.
Practically, a four component spinor index $\tilde{\a}$ of SU*($4$) can be 
replaced by a pair of undotted and dotted indices $(\a,\ad)$ of 
Sl($2,{\mathbb C}$) and a $(1,0)$ Weyl spinor can be written as
\beq
\Psi^{\tilde{\a}}=\(\begin{array}{c} \psi^\a_1\vspace{1ex}\\
\Bar{\psi}^\ad_2 \end{array}\)~~~;~~~ \Bar{\Psi}^{\tilde{\a}}=
\(\begin{array}{c} -\psi^\a_2\vspace{1ex}\\
\Bar{\psi}^\ad_1\end{array}\)
\eeq
where $\Bar{\Psi}^{\tilde{\a}} \equiv {\cal
C}^{\tilde{\a}}_{~~\dot{\tilde{\b}}}\(\Bar{\Psi}\)^{\dot{\tilde{\b}}}$
is the complex--conjugated of
$\Psi^{\tilde{\a}}$ written in the left representation using the 6D
charge--conjugation matrix ${\cal C}^{\tilde{\a}}_{~~\dot{\tilde{\b}}}=
\(\begin{array}{cc}0&-\d^\a_{~\b}
\vspace{1ex}\\\d^\ad_{~\bd}&0\end{array}\)$.

The six--dimensional gamma matrices
$\G^\m$ $\m=0,\dots,5$ acting on $(1,0)$ Weyl spinors can be
represented as
\beq
\G^\m_{\tilde{\a}\tilde{\b}}=\(\begin{array}{cc}\G^\m_{\a\b}& \G^\m_{\a\bd}
\vspace{1ex}\\
-\G^\m_{\b\ad}&\G^\m_{\ad\bd}
\end{array}\)~,
\eeq
with
\bea
&\G^a_{\a\bd} = \s^a_{\a\bd}~~~,~~&\G^a_{\a\b}=\G^a_{\ad\bd}=0~~~,
~~~~(a=0,1,2,3)~~~~;\nonumber \\
&\G^4_{\a\bd} = 0~~~~~~,~~&\G^4_{\a\b} = iC_{\a\b}~~~~~~~,
~~~~~\G^4_{\ad\bd}= iC_{\ad\bd}~~~~~~\,;\,\nonumber\\
&\G^5_{\a\bd} = 0~~~~~~,~~&\G^5_{\a\b} = C_{\a\b}~~~~~~~~,~~~~~
\G^5_{\ad\bd}=-C_{\ad\bd}~~~~~,~
\label{Gamma6d(1,0)}
\eea
$\s^a_{\a\bd}$ being the Pauli matrices and $C_{\a\b}=C_{\ad\bd}=
\(\begin{array}{cc}0& -i\vspace{1ex}\\i & 0\end{array}\)$.

Using the SU*($4$) invariant $\e^{\tilde{\a}\tilde{\b}\tilde{\g}\tilde{\d}}$
($\e^{1234}=\e_{1234}=1$) it is possible to raise and lower pairs of
antisymmetric indices. In particular, the gamma matrices 
$\G^{\m\tilde{\a}\tilde{\b}}$ acting on a $(0,1)$ Weyl spinor 
$\Psi_{\tilde{\a}}$ are given by
\bea
\G^{\m\tilde{\a}\tilde{\b}}=
{1\over 2}\e^{\tilde{\a}\tilde{\b}\tilde{\g}\tilde{\d}}
\G^\m_{\tilde{\g}\tilde{\d}}&,&
\G^\m_{\tilde{\a}\tilde{\b}}=
{1\over 2}\e_{\tilde{\a}\tilde{\b}\tilde{\g}\tilde{\d}}
\G^{\m\tilde{\g}\tilde{\d}}~~.
\label{Gammas}
\eea
These matrices satisfy
\beq
\G^\m_{\tilde{\a}\tilde{\b}}\G^{\n\tilde{\b}\tilde{\g}}+
\G^\n_{\tilde{\a}\tilde{\b}}\G^{\m\tilde{\b}\tilde{\g}}=
-2\eta^{\m\n}\d_{\tilde{\a}}^{\tilde{\g}}~~~,~~~
\G^\m_{\tilde{\a}\tilde{\b}}\G_\m^{\tilde{\g}\tilde{\d}}=
4\d_{\tilde{\a}}^{[\tilde{\g}}\d_{\tilde{\b}}^{\tilde{\d}]}~~.
\eeq
where $\eta_{\mu\nu} = {\rm diag}(-1,1, \cdots, 1)$.

Introducing the spacetime derivatives $\pa_{\a\bd} \equiv \s^a_{\a\bd}
\pa_a $ and $\pa \equiv (\pa_4 - i\pa_5)$, $\pab \equiv (\pa_4 + i\pa_5)$
we have
\beq
\pa_{\tilde{\a}\tilde{\b}}\equiv\G_{\tilde{\a}\tilde{\b}}^\m\pa_\m=
\left(\begin{array}{cc}
iC_{\a\b}\pa&\pa_{\a\bd}\vspace{1ex}\\-\pa_{\b\ad}&iC_{\ad\bd}\pab
\end{array}\right)
~~~,~~~\pa^{\tilde{\a}\tilde{\b}}\equiv\G^{\m\tilde{\a}\tilde{\b}}\pa_\m=
\left(\begin{array}{cc}
-iC^{\a\b}\pab&\pa^{\a\bd}\vspace{1ex}\\-\pa^{\b\ad}&-iC^{\ad\bd}\pa
\end{array}\right)~.~
\label{paOp4dnot}
\eeq

The action which describes the free dynamics of a six--dimensional
$(1,0)$ Weyl spinor is
\beq
\int d^6x
\Big[\Bar{\Psi}^{\tilde{\a}}\G^\m_{\tilde{\a}\tilde{\b}}i\pa_\m
\Psi^{\tilde{\b}}\Big]=
\int d^6x \Big[-\Bar{\psi}_1^\ad
i\pa_{\a\ad}\psi_1^\a-\Bar{\psi}_2^\ad i \pa_{\a\ad} \psi_2^\a
-\psi_2^\a\pa\psi_{1\a}-\Bar{\psi}_2^\ad\pab\Bar{\psi}_{1\ad}\Big]~,~~~
\label{free6d(1,0)}
\eeq
whereas for a $(0,1)$ Weyl spinor we have
\beq
\int d^6x
\Big[\Bar{\Psi}_{\tilde{\a}}\G^{\m\tilde{\a}\tilde{\b}}i\pa_\m
\Psi_{\tilde{\b}}\Big]=
\int d^6x \Big[-\Bar{\psi}_1^\ad
i\pa_{\a\ad}\psi_1^\a -\Bar{\psi}_2^\ad i \pa_{\a\ad} \psi_2^\a
+\psi_2^\a\pab\,\psi_{1\a}+\Bar{\psi}_2^\ad\pa\,\Bar{\psi}_{1\ad}\Big]~.~
\label{free6d(0,1)}
\eeq
Given the structure (\ref{paOp4dnot}) for the 6D spacetime
derivatives the action (\ref{free6d(0,1)}) is simply obtained from
(\ref{free6d(1,0)}) by the exchange $\pa\leftrightarrow -\pab$.

\end{document}